\documentclass[longauth]{aa}  

\usepackage{graphicx}
\usepackage{txfonts}
\usepackage{lscape}

\usepackage[colorlinks=true,allcolors=blue]{hyperref}
\newcommand{\oiii}{[O$\scriptstyle\rm~III$]}	
\newcommand{\ha}{H$\alpha$}	
\newcommand{\snr}{S/N}

\newcommand{\sfr}{M${_\odot}$ yr$^{-1}$}	
\newcommand{\msun}{M${_\odot}$}

\newcommand{\kms}{${\rm km~s^{-1}}$}	
\newcommand{\vout}{$v_{\rm out}$}

\newcommand{\mstar}{M$_\star$}	
	
\newcommand{\jadesdeep}{Deep/HST}

\begin{document}

   \title{JADES: The incidence rate and properties of galactic outflows in low-mass galaxies across $3<z<9$}

\author{
Stefano Carniani
\inst{1}\fnmsep\thanks{stefano.carniani@sns.it}
\and
Giacomo Venturi
\inst{1}
\and
Eleonora Parlanti
\inst{1}
\and
Anna de Graaff
\inst{2} 
\and
Roberto Maiolino
\inst{3,4,5}
\and 
Santiago Arribas  \inst{6}
\and 
Nina Bonaventura \inst{7, 8, 9}
\and 
Kristan Boyett \inst{10,11}
\and
Andrew J.\ Bunker \inst{12}
\and
Alex J. Cameron \inst{12}
\and
Stephane Charlot \inst{13}
\and 
Jacopo Chevallard \inst{12}
\and
Mirko Curti \inst{14, 3, 4}
\and 
Emma Curtis-Lake \inst{15}
\and
Daniel J.\ Eisenstein \inst{16}
\and
Giovanna Giardino \inst{17}
\and
Ryan Hausen \inst{18}
\and 
Nimisha Kumari \inst{19}
\and 
Michael V. Maseda \inst{20}
\and
Erica Nelson \inst{21}
\and
Michele Perna \inst{6}
\and
Hans-Walter Rix \inst{2}
\and
Brant Robertson \inst{22}
\and
Bruno Rodr\'iguez Del Pino \inst{6}
\and
Lester Sandles \inst{3,4}
\and
Jan Scholtz \inst{3,4}
\and
Charlotte Simmonds \inst{3,4}
\and
Renske Smit \inst{23}
\and
Sandro Tacchella \inst{3,4}
\and
Hannah \"Ubler \inst{3,4}
\and 
Christina C. Williams \inst{24}
\and
Chris Willott \inst{25}
\and 
Joris Witstok \inst{3,4}
}
\authorrunning{Carniani et al.}
   \date{}
\institute{
Scuola Normale Superiore, Piazza dei Cavalieri 7, I-56126 Pisa, Italy 
\and    
Max-Planck-Institut f\"ur Astronomie, K\"onigstuhl 17, D-69117, Heidelberg, Germany 
\and 
Kavli Institute for Cosmology, University of Cambridge, Madingley Road, Cambridge, CB3 OHA, UK. 
\and
Cavendish Laboratory - Astrophysics Group, University of Cambridge, 19 JJ Thomson Avenue, Cambridge, CB3 OHE, UK. 
\and
Department of Physics and Astronomy, University College London, Gower Street, London WC1E 6BT, UK 
\and 
Centro de Astrobiolog\'ia (CAB), CSIC–INTA, Cra. de Ajalvir Km.~4, 28850- Torrej\'on de Ardoz, Madrid, Spain 
\and 
Cosmic Dawn Center (DAWN), Copenhagen, Denmark 
\and
Niels Bohr Institute, University of Copenhagen, Jagtvej 128, DK-2200, Copenhagen, Denmark 
\and 
Steward Observatory University of Arizona 933 N. Cherry Avenue Tucson AZ 85721, USA 
\and 
School of Physics, University of Melbourne, Parkville 3010, VIC, Australia 
\and
ARC Centre of Excellence for All Sky Astrophysics in 3 Dimensions (ASTRO 3D), Australia 
\and 
Department of Physics, University of Oxford, Denys Wilkinson Building, Keble Road, Oxford OX1 3RH, UK 
\and
Sorbonne Universit\'e, CNRS, UMR 7095, Institut d'Astrophysique de Paris, 98 bis bd Arago, 75014 Paris, France 
\and
European Southern Observatory, Karl-Schwarzschild-Strasse 2, 85748 Garching, Germany 
\and
Centre for Astrophysics Research, Department of Physics, Astronomy and Mathematics, University of Hertfordshire, Hatfield AL10 9AB, UK 
\and
Center for Astrophysics $|$ Harvard \& Smithsonian, 60 Garden St., Cambridge MA 02138 USA 
\and
ATG Europe for the European Space Agency, ESTEC, Noordwijk, The Netherlands 
\and 
Department of Physics and Astronomy, The Johns Hopkins University, 3400 N. Charles St., Baltimore, MD 21218
\and
AURA for European Space Agency, Space Telescope Science Institute, 3700 San Martin Drive. Baltimore, MD, 21210 
\and
Department of Astronomy, University of Wisconsin-Madison, 475 N. Charter St., Madison, WI 53706 USA  
\and
Department for Astrophysical and Planetary Science, University of Colorado, Boulder, CO 80309, USA 
\and
Department of Astronomy and Astrophysics University of California, Santa Cruz, 1156 High Street, Santa Cruz CA 96054, USA 
\and
Astrophysics Research Institute, Liverpool John Moores University, 146 Brownlow Hill, Liverpool L3 5RF, UK 
\and
NSF’s National Optical-Infrared Astronomy Research Laboratory, 950 North Cherry Avenue, Tucson, AZ 85719, USA 
\and
NRC Herzberg, 5071 West Saanich Rd, Victoria, BC V9E 2E7, Canada 
}

 
  \abstract{
  We investigate the incidence and properties of ionised gas outflows in a sample of 52 galaxies with stellar masses between $10^7$~M$_{\odot}$ and $10^9$~M$_{\odot}$ observed with ultra-deep JWST/NIRSpec MSA spectroscopy as part of the JWST Advanced Deep Extragalactic Survey (JADES). The high-spectral resolution (R2700) NIRSpec observations allowed us to identify for the first time the potential signature of outflows in the rest-frame optical nebular lines in low-mass galaxies at $z>4$. The incidence fraction of ionised outflows, traced by broad components, is about 25--40$\%$, depending on the intensity of the emission lines. The low incidence fraction might be due to both the sensitivity limit and the fact that outflows are not isotropic, but have a limited opening angle, which only results in detection when this is directed toward our line of sight.  Evidence for outflows increases slightly with stellar mass and star formation rate.
  The median velocity and mass-loading factor (i.e. the ratio of the mass outflow rate and star formation rate) of the outflowing ionised gas are  350~\kms\ and $\eta=2.0^{+1.6}_{-1.5}$, respectively. These are 1.5 and 100 times higher than the typical values observed in local dwarf galaxies.  Some of these high-redshift outflows can escape the gravitational potential of the galaxy and dark matter halo and enrich the circumgalactic medium and possibly even the intergalactic medium. Our results indicate that outflows can significantly impact the star formation activity in low-mass galaxies within the first 2 Gyr of the Universe.
  }


   \maketitle
%

\section{Introduction}

One long-standing problem in astrophysics is the process that regulates star formation in galaxies. Only a small fraction ($<10-20\%$) of the baryonic matter in the galaxy halo is currently in the form of stars \citep[e.g.][]{Behroozi:2013, Behroozi:2019}, indicating that galaxies have been relatively inefficient in forming stars from their available gas reservoir across cosmic time. Without a mechanism that balances gas accretion and star formation, available models of galaxy formation incorporating the standard cold dark matter paradigm \citep[e.g.,][]{White:1978} expect the stellar-to-baryon fraction to be well above $50-80\%$, depending on the stellar mass \citep[e.g.,][]{Henriques:2019}. The primary mechanisms that regulate star formation in galaxies are still unclear and debated, however. 
Theoretical models suggest that the energy injected by supernova explosions and the radiation pressure from hot, young stars and active galactic nuclei (AGN) drive the bulk motion of gas at velocities of hundreds of \kms\ on galactic scales \citep[e.g.,][]{Debuhr:2012,Li:2017, Nelson:2019, Mitchell:2020, Pandya:2021}. This fast-outflowing gas is thought to have a significant impact on galaxies by removing or heating the supply of cold gas required for star formation, generating turbulence in the interstellar medium (ISM), shaping galaxy morphology and kinematics, and polluting the circumgalactic medium (CGM) with metal-enriched material. Galactic outflows and their feedback processes are required in cosmological simulations to self-regulate the star formation in low-mass and massive galaxies \citep[e.g.,][]{Ceverino:2018}. Thus, we need to study outflow demographics across cosmic time and the impact of feedback on the host galaxy and the surrounding environment to confirm or disprove the current models of galaxy formation and evolution.

In recent years, spectroscopic observations of the strongest rest-frame optical nebular lines, such as \ha\ and \oiii, have provided the most extensive census of galactic outflows from $z = 0$ to $z\sim3$, the so-called ``cosmic noon'' \citep{Arribas:2014, Forster-Schreiber:2014, Harrison:2016, Cicone:2016, Concas:2017,  Rakshit:2018, Perna:2017, Leung:2019, Forster:2019, Davies:2019, Kakkad:2020, Reichardt:2022, Concas:2022, Llerena:2023, Rodriguez:2023}. The considerably high incidence of outflows ($>30\%$, depending on the survey) suggests that outflows are common in massive ($>10^{10}$\msun) galaxies up to cosmic noon \citep[e.g.,][]{Carniani:2015, Cicone:2016, Concas:2017, Rakshit:2018, Forster:2019}. In contrast, the investigation of the outflow properties in the low-mass ($<10^{10}$~\msun) regime is limited to the local Universe because the sensitivity of current ground-based facilities in the near-IR band is not sufficient to detect the faint broad component of nebular lines tracing ionised outflowing gas.

With the advent of the James Webb Space Telescope \citep[JWST;][]{Gardner:2006, Gardner:2023}, we are now able to extend the outflow studies towards higher redshifts and investigate the impact of feedback in galaxies in the first few billion years of cosmic time. In particular, the spectral coverage of the Near Infrared Spectrograph \citep[NIRSpec;][]{Ferruit:2022, Jakobsen:2022} enables the detection of the strongest rest-frame optical nebular lines out to $z\sim9$, and its high sensitivity facilitates the identification of slow and weak galactic outflows in low-mass galaxies that separate them from large-scale orbital motions of the host galaxy. For example, \cite{Tang:2023}, \cite{Zhang:2023}, and \cite{Xu:2023} reported the detection of a broad emission-line component that likely traces ionised outflows in seven targets of the Cosmic Evolution Early Release Science \citep[CEERS;][]{Bagley:2023, Finkelstein:2023}, Early Release Observations \citep[ERO;][]{Pontoppidan:2022}, GLASS JWST Early Release Science \citep[GLASS-JWST-ERS;][]{Treu:2022} surveys, and part of the public data from the JADES surveys \citealt{Eisenstein:2023, Rieke:2023}. 

In this work, we study the incidence of outflows in the NIRSpec GOODS-S Deep/HST pointing (Program ID: 1210; PI: N. Luetzgendorf) of the JWST Advanced Deep Extragalactic Survey (JADES; \citealt{Eisenstein:2023, Rieke:2023} ).  The NIRSpec Deep/HST pointing provides R100, R1000, and R2700 spectroscopy of 253 galaxies in the Hubble Ultra Deep Field and surrounding GOODS-South \citep{Bunker:2023}. We specifically exploit the deepest R2700 observations carried out with the spectral configuration G395H/F290LP, which covers the wavelength range between 2.87~$
\mu$m  and 5.14~$\mu$m and allows us to detect the strongest rest-frame optical nebular lines up to $z\sim9$. The high sensitivity of these observations (i.e. 5$\times10^{-19}~{\rm erg~s^{-1}~cm^{2}}$ for an unresolved emission line at $4~\mu$m) allows us for the first time to analyse the main properties of ionised outflows in a sample of low-mass ($<10^{10}$~\msun) galaxies in the first 2 Gyr of cosmic time ($3<z<9$).

The paper is organised as follows. Th sample selection and data reduction are presented in Sec.~\ref{sec:sample} and Sec.~\ref{sec:datareduction}, and the outflow identification workflow is described in Sec.~\ref{sec:outflow_identification}. We analyse the incidence and properties of ionised outflows in Sec.~\ref{sec:incidence} and Sec.~\ref{sec:properties}, respectively. Finally, we summarise our conclusions in Sec.~\ref{sec:conclusions}.
We adopt the following cosmological parameters from \cite{Planck-Collaboration:2015}: $H_0$ = 67.7 km s$^{-1}$ Mpc$^{-1}$, $\Omega_{\rm m}$ = 0.308 and  $\Omega_{\rm \Lambda}$ = 0.70, according to which 1\arcsec\ at $z = 6$ corresponds to a  physical scale of 5.84~kpc. Hereafter, we use the terms low-mass and high-mass galaxy to refer to systems with a stellar mass $M_\star\lesssim10^{9}~{\rm M_{\odot}}$ and $M_\star\gtrsim10^{9}~{\rm M_{\odot}}$, respectively.  Throughout
this work, we use the term SFR to refer to the star formation rate averaged over the past 10 Myr.

\section{Sample selection}
\label{sec:sample}

   \begin{figure}
   \centering
   \includegraphics[width=0.5\textwidth]{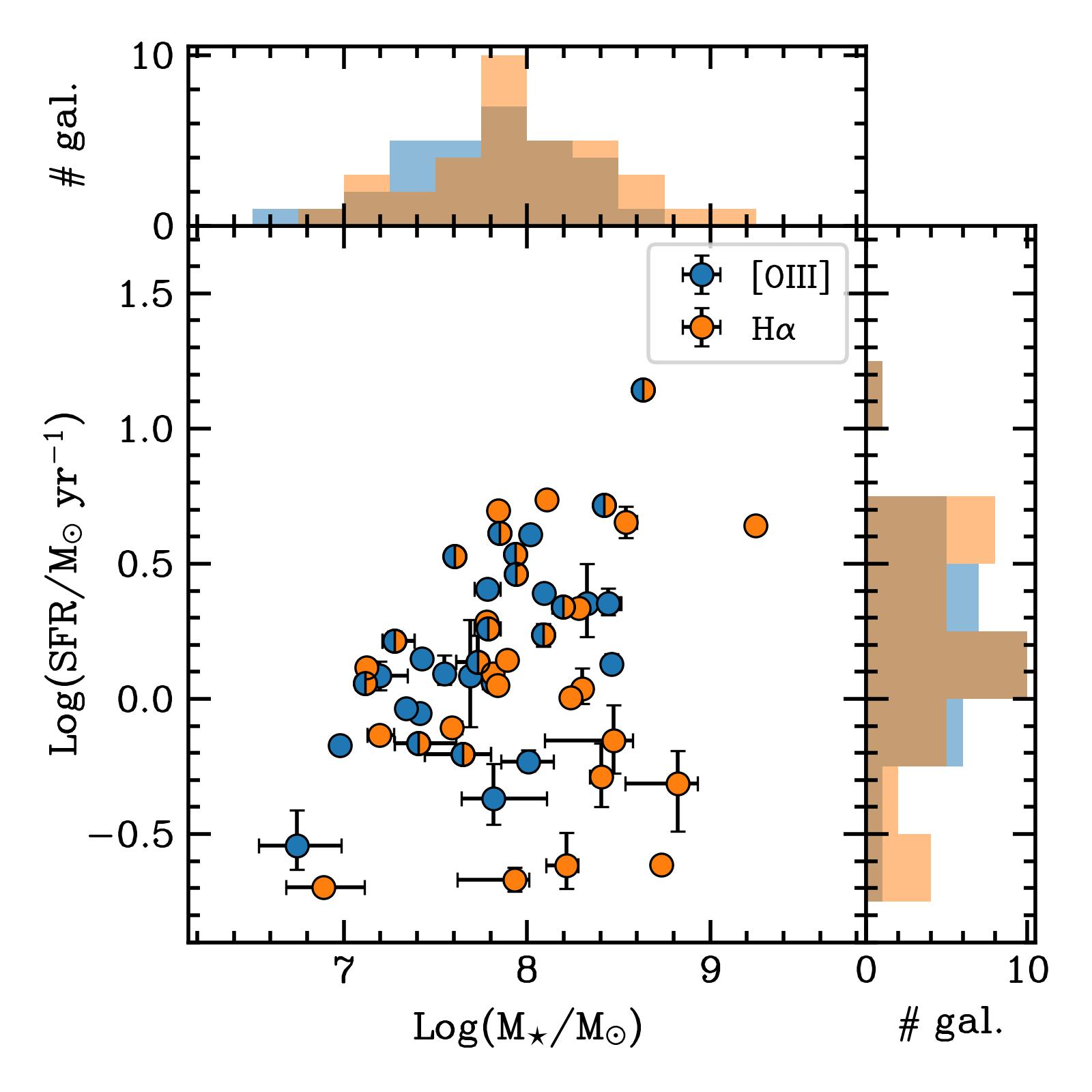}

   \caption{Star formation rate vs. stellar mass ($M_\star$) for the 52 JADES galaxies at $3.5<z<8.5$. The selected galaxies are detected in \oiii, \ha, or in the NIRSpec G395H and F290LP bands with a \snr~$>$ 5. Targets detected in \oiii\ and \ha\ are reported in blue and orange, respectively. Galaxies detected in both \oiii\ and \ha\ are illustrated with circles. Half of the marks are shown in blue and the other half in orange.  The right and upper panels show the number of targets in bins of SFR and $M_\star$, respectively.}
              \label{fig:sample}%
    \end{figure}

We selected galaxies detected in either \oiii\ or \ha, or in the NIRSpec G395H and F290LP 1D spectra with a peak signal-to-noise (\snr) ratio higher than 5. We defined the \snr\ of an emission line as the ratio of the flux peak of the line and the sensitivity level of the observation at the same wavelength.

Our sample comprises 52 JADES galaxies that span redshifts between 3.5 and 8.5 and the following ranges of stellar masses and SFRs, the latter averaged over the past 10 Myr:  $\log(M_\star/{\rm M_\odot})\in[6.7-9.2]$, SFR $\in[0.32~{-~28.2}]~\rm M_\odot~yr^{-1}$.  Stellar masses and SFRs were estimated from the NIRSpec PRISM/CLEAR spectra and Kron-based photometry \citep{Rieke:2023} with the \textsc{BEAGLE} code \citep{Chevallard:2006}. In the fitting, we used the synthetic stellar population described in \cite{Vidal-Garcia:2017} and adopted a delayed-exponential profile for the star formation history and a Chabrier initial mass function \cite{Chabrier:2003}. 
A low-order calibration polynomial was included in the fitting procedure to match the shape and normalisation of the spectrum to the photometry. This allowed us to correct for any additional slit losses that were not fixed in the NIRSpec GTO pipeline.
A detailed description of the fitting procedure and all the free parameters is given in Chevallard et al., in prep. 

  The sample distribution in the SFR versus stellar mass plane is shown in Figure~\ref{fig:sample}. We stress that ionised outflows can be identified and studied in individual galaxies in the low-stellar mass regime ($M_\star\lesssim10^{9}~ {\rm M_\odot}$) at $z>3-4$ by using the rest-frame optical lines for the first time.

\section{NIRSpec observations}
\label{sec:datareduction}

As discussed in \cite{Bunker:2023}, in the NIRSpec Deep/HST program, we assigned three shutters aligned along the spatial direction to form a slitlet to each target. The telescope was also nodded by one shutter along the spatial direction between each nodding, such that the targets were observed in each shutter in a sequence. This method enables the pixel-to-pixel subtraction at the count rate level. All targets were observed with the low-resolution configuration (PRISM), while some bright low-priority and low-\textit{z} galaxies were removed from the high spectral resolution (R1000 and R2700) observations.

We used the flux-calibrated 1D data produced using a custom pipeline developed by the
ESA NIRSpec Science Operations Team (SOT) and Guaranteed Time Observations (GTO) teams. Most of the pipeline workflow adopts the same algorithms as included in the official STScI pipeline \citep{Alves-de-Oliveira:2018, Ferruit:2022}. While the details of this custom pipeline will be presented in a forthcoming paper (Carniani et al., in prep.), the main steps are discussed in \cite{Curtis-Lake:2023} and \cite{Bunker:2023}.

Differently from the standard products delivered by the pipeline, we used a finer regular grid wavelength to improve the sampling of the line spread function in the 2D rectification process. 
We adopted a regular grid wavelength with a pixel scale of 3.32~\AA  so that the line spread function was sampled by at least four spectral channels. 
This oversampling is adequate to map the optical nebular line profile and improve the detection of broad components tracing fast-emitting gas. We also used the 1D spectra extracted from a three-pixel aperture to maximise the signal-to-noise ratio of the signal.

\section{Outflow identification}
\label{sec:outflow_identification}

To investigate potential outflow signatures in the 1D spectra of the galaxies in our sample, we performed a multi-Gaussian fitting of the emission lines using the nested sampling
algorithm \textsc{dynesty} \citep{Speagle:2020} and disentangled the emission associated with the galactic disk from the emission associated with fast gas. In this section, we describe the spectral fitting and outflow identification process in detail.

Initially, we estimated the continuum emission due to either stellar populations or potential residuals after background subtraction around the wavelengths of the nebular lines. In particular, we determined the continuum level in a velocity range of $1000~{\rm km~s^{-1}}<|v|<2000~{\rm km~s^{-1}}$ from the peak of the emission line to exclude the spectral range covered by the line and possible broad components due to outflows. 
We note that the continuum emission of the R2700 data is only detectable in the brightest galaxies of the sample. We thus fitted the continuum underlying the emission lines by using a constant and neglected the contribution of stellar absorption to Balmer lines.
We then performed two fits of the continuum-subtracted spectrum: the first fit with a single Gaussian component, and the second fit with a double Gaussian model, defined as
$$
   \begin{cases}
   F_\lambda(\lambda) = A_{\rm narrow}\exp\left(-\frac{1}{2}\frac{(\lambda-\lambda_{\rm narrow})^2}{\sigma_{\rm narrow}^2}\right)+A_{\rm broad}\exp\left(-\frac{1}{2}\frac{(\lambda-\lambda_{\rm broad})^2}{\sigma_{\rm broad}^2}\right),\\
 \sigma_{\rm broad} = \sigma_{\rm narrow}\delta_\sigma, \\
  A_{\rm broad} = A_{\rm narrow}\delta_A,
    \end{cases}
$$
where $F_\lambda(\lambda)$ is the emission line profile as a function of wavelength, $A_{\rm narrow}$, $\lambda_{\rm narrow}$, and $\sigma_{narrow}$ are the amplitude, centroid, and velocity dispersion, respectively, of the narrow Gaussian component associated with the galactic disk, and $A_{\rm broad}$, $\lambda_{\rm broad}$, and $\sigma_{\rm broad}$ are the same parameters as for the broad Gaussian component tracing the outflowing gas. 
The free parameters for the broad component are $\delta_A$, $\lambda_{\rm broad}$, and $\delta_\sigma$.
We imposed $\sigma_{\rm broad} = \sigma_{\rm narrow}\delta_\sigma$ with $\delta_\sigma>1.2$ so that the velocity dispersion of the additional component was at least 20\% higher than that of the first component, without setting any fixed lower limit for $\sigma_{\rm broad}$ for each galaxy. A similar approach was adopted for the amplitude parameter, defined as $A_{\rm broad} = A_{\rm narrow}\delta_A$ with $\delta_A<0.5$ based on low-redshift outflow studies \citep{Swinbank:2019, Concas:2022}.
We note that these constraints allowed us to automatically exclude unphysical best-fitting results in which the emission line is reproduced by two components with a similar velocity dispersion and amplitude. In these cases, the component associated with the host galaxy would be too narrow ($\sigma<40-60$~\kms\ depending on the stellar mass) to represent a spatially integrated emission line arising from either a rotation- or pressure-supported disk with the mass and size of our targets.


For all free components, we used a flat distribution for the priors.
We only set an upper bound of $150~{\rm km~s^{-1}}$  on the priors of $\sigma_{narrow}$, which corresponds to the maximum broadening that the narrow component can reach based on a) the broadening due to the rotational motions in the disk given the mass and size of each galaxy (see~App.~\ref{app:best_fitting_results}), b) the velocity dispersion of gas observed in $z>4$ galaxies, which spans a range between 30~\kms\ and 80~\kms \citep{Lelli:2021, Lelli:2023, Parlanti:2023, Rizzo:2023}, and c) the line spread function of the instrument that enlarges the line width of the emission line depending on the intra-shutter position and size of the target \citep{Jakobsen:2022,de_Graaff:2023}.

The presence of outflows in each galaxy was determined based on the presence of a second broad component in addition to the narrow component associated with the general system motions. In turn, the two-Gaussian best-fit model was preferred to the one-Gaussian one based on the following criteria (to be satisfied jointly):  
\begin{itemize}
    \item the two-Gaussian model led to a decrease in reduced chi-squared ($\chi^2_r$) by 10\% or more compared to $\chi^2_r$ of the single-component model;  
    \item the log-difference between the two Bayesian evidences\footnote{Bayesian evidence is the average of the likelihood under the prior for a specific model choice and is used to quantify the support for one model over the other \citep{Skilling:2004}. 
    A value of $\Delta\log K = \log K_{\rm double} -  \log K_{\rm single} > 2$, where $K_{\rm double}$ and $K_{\rm single}$ are the Bayesian evidence estimated for the double- and single-Gaussian models, respectively,  means that the former model is more strongly supported by the data under consideration than the latter.} ($\Delta\log K$) of the models was larger than 2 \citep{Jeffreys:1961};
    \item integrated flux emission of the broad component was higher than three times the uncertainties.
\end{itemize}
We also performed a visual inspection to verify that the best-fitting broad component model was not associated with noise fluctuations or continuum that had not been fully or properly subtracted. 

From the targets with candidate outflows detected in $H\alpha$, selected as described above, we removed two targets, 10013704 and 8083, because the broad component of \ha\ was not associated with outflowing gas, but arose from the broad-line region of AGN based on the analysis of the nebular lines in R1000 NIRSpec data (\citealt{Bunker:2023} and \citealt{Maiolino:2023b}).  On the other hand, we note that 10013704 shows evidence for outflows in \oiii\, and this target was therefore included in the \oiii\ outflow sample. After applying these selection criteria, we identified 13 galaxies with outflow signatures. The spectra of all of them, together with their best-fitting models, are reported in Appendix~\ref{app:best_fitting_results}.  The best-fit results for the modelling with two Gaussian profiles are listed in Tables~\ref{tab:table1} and ~\ref{tab:table1b}.

Although low-redshift ionised outflows are often observed as a blueshifted broad component, the identified broad components in our sample are both blue- and redshifted relative to the systemic velocity of the galaxy. The lack of detected broad redshifted components in low-redshift galaxies is mainly associated with the absorption along our line of sight of the emission from the receding outflowing gas extended or located beyond the galaxy due to dust in the interstellar medium. However, our sample displays a  modest dust content with a median dust attenuation value of $A_{\rm v}\sim0.4$ mag, and a broad blue- or redshifted component therefore mainly depends on the outflow morphology. Simulations \citep[e.g.,][]{Nelson:2019} and local observations \citep[e.g.,][]{Venturi:2018, Mingozzi:2019, Husemann:2019, Lopez-Coba:2020, McPherson:2023} both rarely reveal symmetric biconical outflows in edge-on galaxy where the effect of dust absorption along the line of sight is limited. Therefore, in the absence of strong obscuration by the interstellar medium, the outflowing gas might appear as either blue- or redshifted broad profiles.

We disfavour the inflow scenario as an interpretation of the broad components because inflowing gas is expected to have velocities lower than the virial velocity of the halo \citep{Goerdt:2015}, which corresponds to about $80-90$~${\rm km~s^{-1}}$\footnote{The virial velocity is defined as $v_{\rm vir}=\sqrt{GM_{\rm DM}/r_{\rm vir}}$, where $M_{\rm DM}$ and $r_{\rm vir }$ are the virial mass and radius of the dark matter halo, respectively. When we assume a galaxy with $M_\star = 10^{8}~{\rm M_\odot}$, which is expected to form in a dark matter halo of $M_{\rm DM}=3\times10^{10}~{\rm M_\odot}$ with a  $r_{\rm vir }=20~{\rm kpc}$ at $z=6$ \citep[][e.g.,]{Barkana:2001}, we infer $v_{\rm vir}\sim90$~\kms.}  for the galaxies in our sample. We thus expect that filaments of inflowing cold gas with low velocities like this yield an emission line width that is smaller than those observed in our sample ($<\sigma_{\rm broad}>$=170~\kms). However, we cannot exclude a priori that a fraction of the light of the broad component is associated with inflowing gas.

We note that \cite{Marasco:2023}, who studied the ionised outflows in a sample of local (distance $<400$~Mpc) dwarf ($M_\star = 10^{6}-10^{10}~{\rm M_\odot}$) galaxies, have proposed two outflow scenarios to interpret the broad components. In the first scenario, the feedback mechanism increases the turbulence of the gas in the galaxy without affecting its bulk motion, and the outflows are only composed of gas with radial velocities higher than the escape velocity. In this case, the outflow component is only given by the wings of the broad component at high velocity (higher than the escape velocity\footnote{In \cite{Marasco:2023}, the escape velocity is defined as the minimum speed required to reach the virial radius. We used the traditional definition for the escape velocity:  The minimum speed required to reach an infinite distance.}), while the bulk of the line traces turbulent gas. In the second scenario, the feedback mechanism does not affect the gas turbulence, but only its bulk motion, and the broad component is due to the superposition of the different contributions to the line-of-sight velocity of each outflowing gas direction. Based on their data, the authors suggested that the second scenario is less plausible because a spherically expanding shell of gas with the same velocity in all directions would lead to a boxcar-shaped emission line and not to a Gaussian profile, as observed. However, their study was only based on local spatially resolved observations with a spatial resolution of about 10 parsecs, where the broad wings are only visible in the spectra extracted from the individual pixels. In the spatially integrated emission, the shape of the ionised line is dominated by the narrow component, and it is not possible to determine the profile of the broad components. In our case, the broad component is well visible in the spatially integrated emission line and traces gas on kiloparsec scales.  In this case, a spherical expanding shell, which may apply for an SN-driven bubble on scales of tens of parsec, would not be a representative scenario to reproduce the outflows on galactic scales. Simulations indeed predict that gas accelerated by superimposed spherically expanding shells follows the path of least resistance through the interstellar medium and results in bipolar outflows with a wide opening angle on large scales \citep{Cooper:2007, Nelson:2019, Schneider:2020}.  Therefore, a conical (or biconical) morphology is more reliable to reproduce the emission of the fast gas, and \cite{Bae:2016} and \cite{Marconcini:2023}  showed that a conical outflow with a constant radial velocity produces a broad component in the line-of-sight integrated emission line profile, which is similar to what is observed in JWST JADES spectra.  We therefore assumed that the whole broad component traces outflowing gas that is moving outward with a constant velocity.

\section{Outflow demography}
\label{sec:incidence}

\subsection{Outflow incidence}

\begin{figure}
   \centering
   \includegraphics[width=0.45\textwidth]{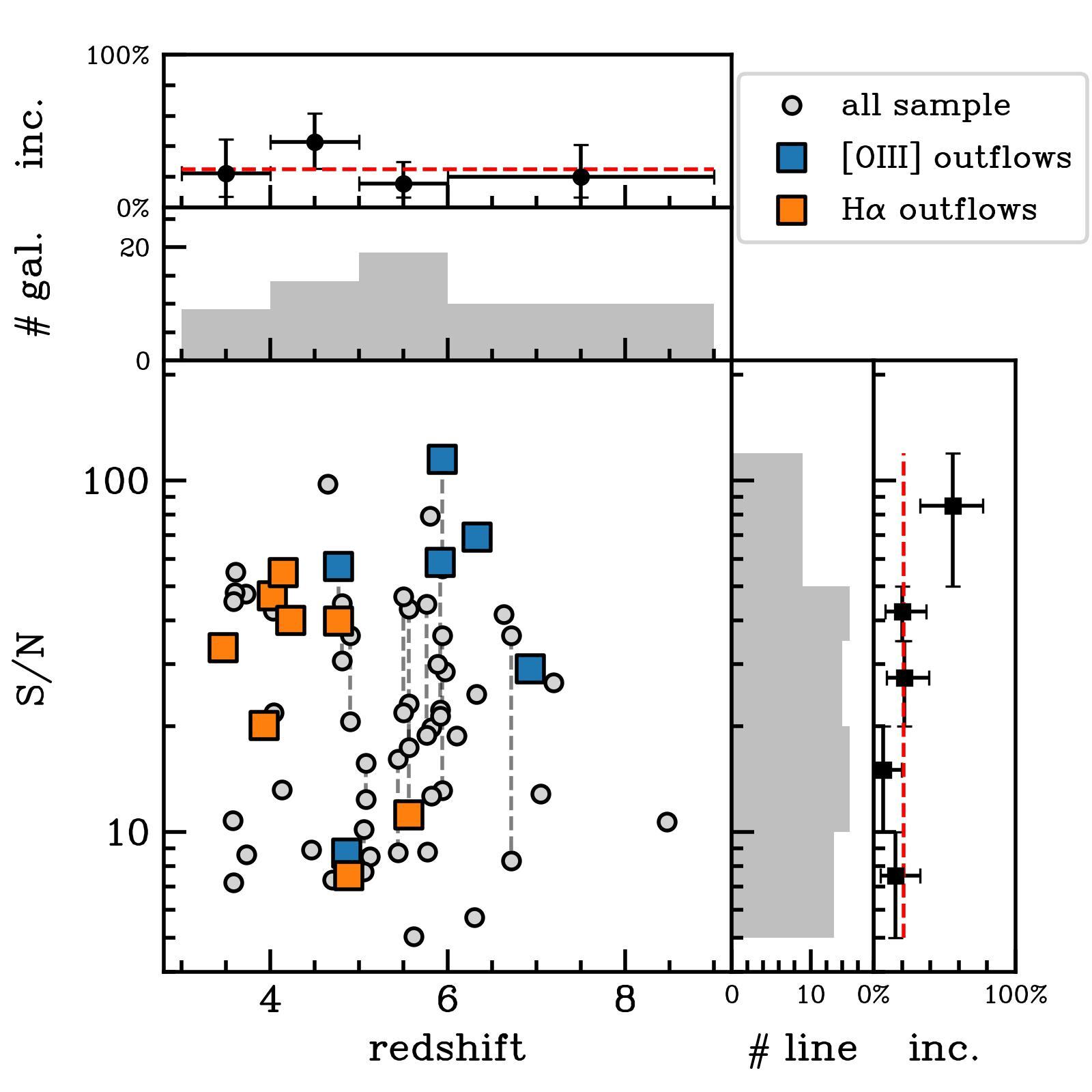}

   \caption{\snr\ of the overall nebular line profiles as a function of redshift. All the detected lines in our sample are reported as grey circles, while the emission lines with evidence of the broad component are indicated with blue (\oiii) and orange (\ha) squares. The dashed grey lines indicate \oiii\ and \ha\ lines observed in the same galaxy. The top panels show the distribution of galaxies and the incidence of outflow in regular redshift bins. The right panels report the distribution of emission lines with S/N $>5$ and the incidence of broad components in irregular bins of \snr, respectively. Each irregular bin of \snr\ contains a similar number of lines, except for the last bin at the highest S/N values. The dashed red line marks an incidence of 25$\%$.}
              \label{fig:snr_vs_z}%
    \end{figure}

Thirteen of the 52 galaxies selected from the JADES survey of the \jadesdeep\ field show outflow signatures that satisfy the requirements described in the previous section. They are reported in Figure~\ref {fig:snr_vs_z}, which shows the \snr\ of the emission lines as a function of redshift. Out of the 13 outflows identified from broad components in the emission line profiles, 5 were only detected in \oiii\ and 7 only in \ha. Only one subsample of the 14 galaxies that were observed in both lines reveals evidence for broad components in both \ha\ and \oiii. Using the two nebular lines, we derived an outflow incidence rate of $\sim25\%$ that appears constant within the uncertainties over the redshift range covered by our sample (top panel of Figure~\ref {fig:snr_vs_z}).

\begin{figure}
   \centering
   \includegraphics[width=0.45\textwidth]{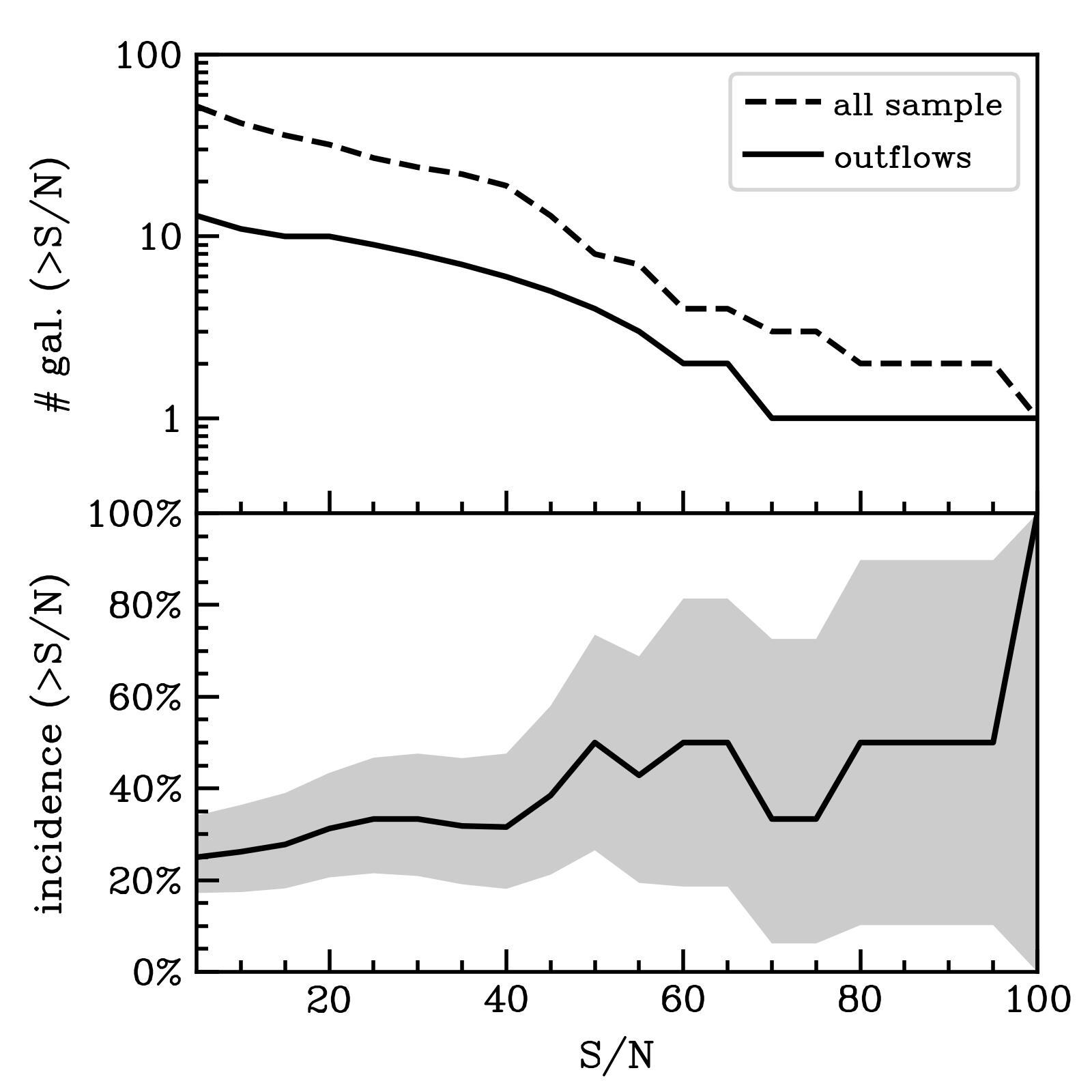}

   \caption{Outflow detection rate vs. S/N of the overall line profile. The cumulative number of targets and detected outflows is reported in the top panel, and the incidence rate is illustrated in the bottom panel. }
    \label{fig:snr}%
    \end{figure}

However, focussing on the fraction of outflow detections as a function of the \snr\ of the overall nebular line (right panel of Figure~\ref {fig:snr_vs_z}), we note that the incidence drops below $25\%$ when we only select galaxies whose nebular lines are detected with 5<\snr<20. Since the targets with \snr<20 represent half of the sample, the low incidence rate could be affected by these fainter galaxies whose sensitivity is not sufficient to detect the broad outflow components in the emission line profiles.  
The dependence of the outflow incidence on the \snr\ is illustrated in Figure~\ref{fig:snr}. The incidence reaches a value of $0.5\pm0.2$ when only targets with \snr$>50$ are selected. However, the number of targets above this threshold is limited ($<9$), and the uncertainties on the incidence fraction are $>50\%$, yielding inconclusive results.

A detection fraction of $25-40\%$ may be caused by the geometry of the outflowing gas.
When we assume a biconical outflow model with an opening angle of $\theta=45\deg$, which is consistent with what is typically measured in the local Universe \citep[e.g.,][]{Venturi:2018, Lopez-Coba:2020, Juneau:2022, Kakkad:2023}, the solid angle subtended by the outflow is
$\Omega = 2\times(2\pi(1-\cos\theta)) \approx 3.68$ steradians,
where the factor of 2 is due to the biconical morphology.
By taking into account that the radiation emitted by outflowing gas in spatially integrated spectra can be clearly distinguished from the radiation from the galactic disk alone when the fast gas moves along the line of sight, particularly in the case of low outflow velocities \citep[e.g.,][]{Woo:2016, Lamperti:2022}, the probability of detecting the broad component can be approximated to $\Omega/4\pi\approx 0.3$, which is consistent with the incidence rate indicated by our observations. Under these assumptions, the inferred outflow incidence may indicate that outflows are common and even ubiquitous in the low-mass galaxy population at $z>3$.

\subsection{Host galaxy properties}
As discussed in the previous section, we find a global outflow detection rate of $25-40\%$ depending on the \snr\ threshold. This indicates that it is more likely to detect a broad component tracing outflows when the \snr\ of the emission line is higher. On the other hand, the incidence fraction may also depend on the galaxy properties.
Previous studies at lower redshift have concluded that the incidence fraction increases with increasing stellar mass \citep{Weiner:2009, Genzel:2014,Forster:2019, Leung:2019, Concas:2022}. Specifically, these studies found an incidence fraction $<10\%$ in galaxies with $M_{\star}\sim10^{10}~{\rm M_{\odot}}$, while the fraction reaches values of $\sim60\%$ in massive galaxies with $M_{\star}\sim10^{11}~{\rm M_{\odot}}$. This trend was associated with the presence of AGN, which are more luminous in massive galaxies, which means that the outflows they accelerate are more likely to be detected because they are faster and more massive than in lower-mass galaxies. We therefore investigated whether the outflows in our sample might also depend on the galaxy properties. However, the sample of galaxies without AGN signatures shows that the incidence of outflows increases with the SFR, supporting the scenario in which this fast gas is accelerated by star formation activity.

\begin{figure*}
   \centering
   \includegraphics[width=1\textwidth]{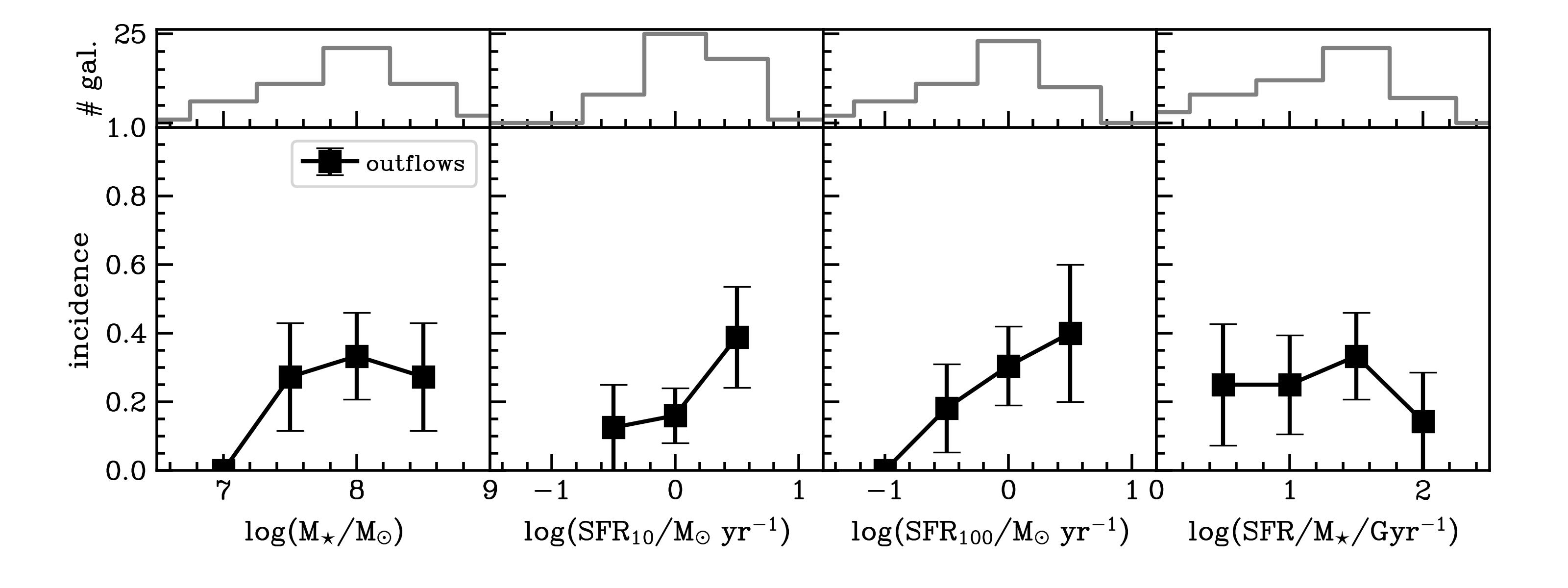}

   \caption{Incidence of outflows as a function of host galaxy properties.  The bottom panels from left to right report the detection rate as a function of stellar mass, star formation rate estimated over the last 10~Myr, and specific star formation rate. The binomial uncertainties are reported as error bars for each bin. The uncertainties on the galaxy properties are smaller than the bin sizes. The top panels illustrate the number of targets in each bin. Incidence values are not reported for bins with fewer than five targets.}
\label{fig:incidence_vs_properties}%
    \end{figure*}

Figure~\ref{fig:incidence_vs_properties} shows the incidence of broad components (i.e. outflows) in the galaxies of the sample as a function of stellar mass, star formation rate, and specific star formation rate (i.e. SFR/\mstar).  We find a positive correlation between the incidence rate and both \mstar\ and SFR. This may be interpreted as evidence that massive and actively star-forming galaxies drive outflows more frequently. However, it is important to stress that the \snr\ of the nebular lines in our sample increases with increasing SFR (Appendix~\ref{sec:selection_bias}), and the trend between incidence rate and SFR might therefore just reflect the correlation between the outflow detection rate and S/N shown in Figure~\ref{fig:snr}.  On the other hand, the stellar mass in Deep/HST pointing does not reveal any correlation with the strength of the nebular lines (Appendix~\ref{sec:selection_bias}), and the positive trend between incidence rate and stellar mass does not seem to be a consequence of selection effects and probably has a physical origin.  
%
%
The positive correlation between outflow incidence and stellar mass might indicate that outflow velocities and kinetic energies are boosted by the radiation from faint AGN that are not detected with standard diagnostics (e.g. [OIII]/H$\beta$ versus [NII]H$\alpha$ diagram; see \citealt{Maiolino:2023b} and \citealt{Scholtz:2023b}). This scenario was recently presented in several theoretical studies \citep{Silk:2017, Dashyan:2018, Koudmani:2019} and is supported by the growing number of works revealing AGN in low-redshift dwarf (\mstar~$<10^{10}$~M$_{\odot}$) galaxies \citep[e.g.,][]{Sartori:2015, Penny:2018, Reines:2020, Mezcua:2023}. The boost due to AGN radiation in more massive galaxies might increase the chance of detecting the broad component in nebular lines.

To reduce the dependence of the outflow incidence on the S/N of the nebular line, we report in the third panel of Figure~\ref{fig:snr} the outflow demography as a function of the SFR averaged over the last 100 Myr estimated by \textsc{beagle} (SFR$_{\rm 100}$). This parameter is less strongly correlated to the strength of the nebular line (Appendix~\ref{sec:selection_bias}), and the positive trend between outflow incidence and  SFR$_{\rm 100}$ suggests that these outflows are mainly driven by star formation activity. However, improving the statistical robustness of this sample is fundamental for confirming these trends and identifying the mechanism that drives outflowing gas in the early Universe.

The last panel of Figure~\ref{fig:incidence_vs_properties} illustrates the outflow incidence rate as a function of the specific star formation rate.  Over the range covered by our sample ($10~{\rm Gyr^{-1}}<{\rm SFR/M_{\star}}<100~{\rm Gyr^{-1}}$), the outflow incidence is constant within the error over the range of the specific star formation rate, but the low statistical significance does not allow us to identify a trend between the two observables. The detection of  outflows in our sample is generally consistent with the framework in which  galaxies whose specific star formation rate exceeds  $\sim10~{\rm Gyr^{-1}}$ might develop radiation-pressure-driven  outflows that clear the galaxy
of its gas and dust \citep{Ziparo:2022, Fiore:2023}.

\section{Outflow properties}
\label{sec:properties}
In this section, we estimate the physical properties of the
outflows and explore the trends with galaxy properties.

\subsection{Outflow velocity}

We first focus on the velocities of the outflowing gas (\vout), which we obtained with the prescription from \cite{Rupke:2005}, who defined the velocity of the outflowing gas as 
\begin{equation}
v_{\rm out} =  |v_{\rm broad} - v_{\rm narrow}| + 2\sigma_{\rm broad,deconv},
\end{equation}

where  $|v_{\rm broad} - v_{\rm narrow}|$ is the shift between the peak velocities of the broad and narrow components (the latter is assumed to trace the systemic velocity), and $\sigma_{\rm broad,deconv}$ is the velocity dispersion of the broad component, deconvolved by the instrumental line spread function $\sigma_{\rm LSF}$ as $\sigma_{\rm broad,deconv}^2 = \sigma_{\rm broad}^2 - \sigma_{\rm LSF}^2$. In any case, the correction is small since $\sigma_{\rm LSF}^2 \sim 40-60$ km/s (depending on wavelength), and the measured $\sigma_{\rm broad}$ are in the range $120-600$ km/s.

This method guarantees that the outflow velocity estimates do not depend strongly on the inclination of the outflow cone with respect to the line of sight \citep{Rupke:2005, Fiore:2017}. 
The outflow velocity estimates cover a range between 300~\kms\ and 1200~\kms, and the individual values for each galaxy are reported in Tables~\ref{tab:table1} and ~\ref{tab:table1b}. 

\begin{figure}
   \centering
   \includegraphics[width=0.5\textwidth]{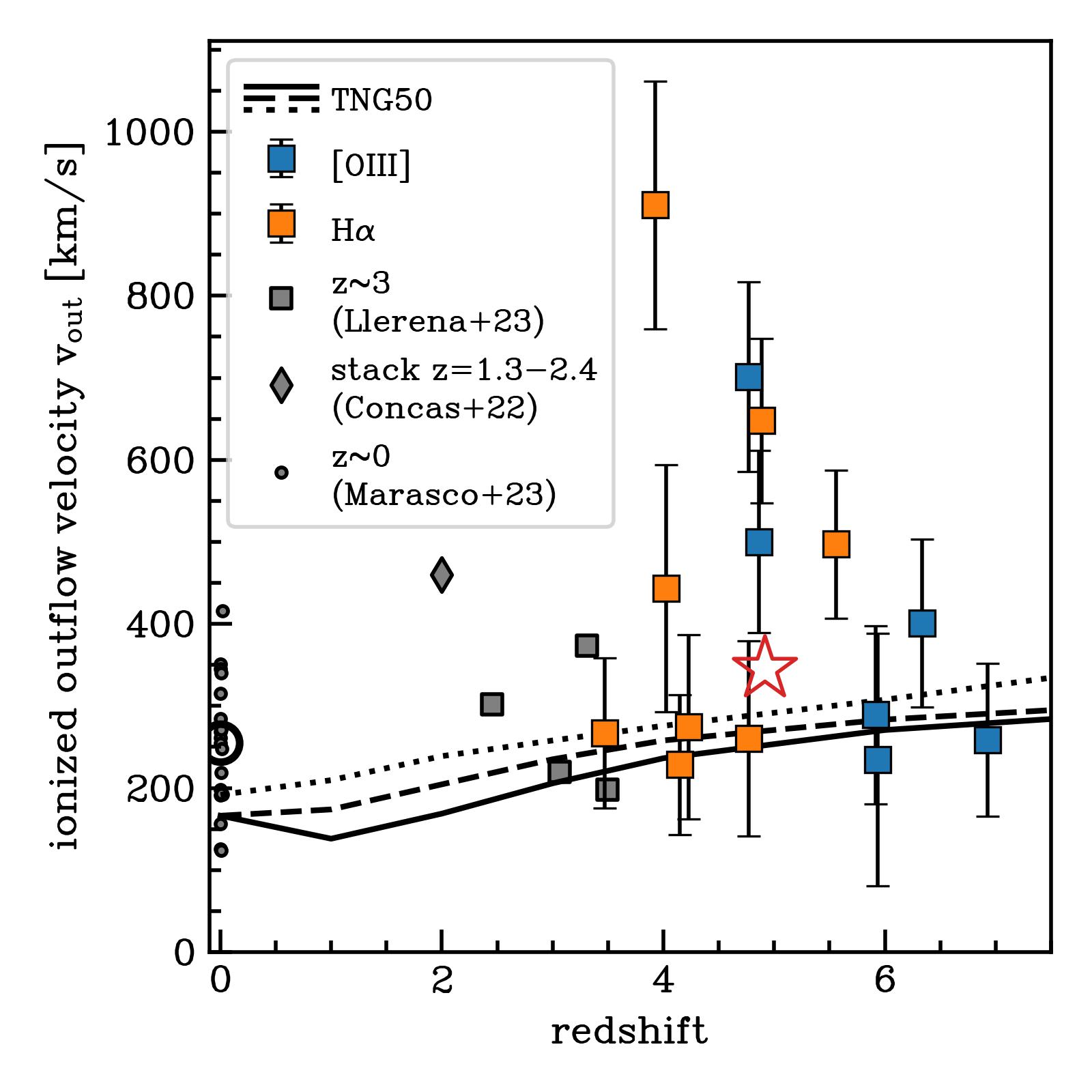}

   \caption{Cosmic evolution of ionised outflow velocities from rest-frame optical emission lines in low-mass ($ M_\star<10^9~{\rm M_\odot}$ ) galaxies. The blue and orange squares indicate the outflows from JADES. The red star represents the median outflow velocity of our sample. The grey circles are the outflow velocities from a sample of local low-mass galaxies \citep{Marasco:2023}  and re-estimated consistently with our definition (see details in the text). The larger empty circle is their median.  The diamond shows a tentative outflow detection by stacking spectra of low-mass galaxies at $z=1.2-2.6$ \citep{Concas:2022}.  The squares report the outflow velocities detected in four low-mass galaxies at $z\sim3$  \citep{Llerena:2023}.  The dotted, dashed, and solid black lines are the predictions from the TNG50 simulation for three different stellar masses: $M_{\rm \star}=10^{8.5},10^{8}, and10^{7.5}~{\rm M_{\odot}}$ \citep{Nelson:2019}.}
              \label{fig:vout_vs_z}%
    \end{figure}

The evolution of outflow velocity as a function of cosmic time for low-mass galaxies is reported in Figure~\ref{fig:vout_vs_z}, where we compare the values found for our sample with those observed at lower redshifts. We stress that ionised outflow studies that are based on rest-frame optical nebular lines in low-mass galaxies are limited to a few samples because the sensitivity of ground-based observations is limited. 
In the local Universe, \cite{Marasco:2023} investigated outflow properties in a sample of 19 nearby galaxies with stellar masses $10^7~{\rm M_{\odot}} < M_\star < 10^{10}~{\rm M_{\odot}}$, mostly lying above the main sequence of star-forming galaxies.
Since the authors assumed the outflow velocity to be equal to the galaxy escape speed, we re-estimated their outflow velocities by calculating the median of the full width at half maximum (FWHM) distribution obtained from the pixel-by-pixel line fitting of the broad components and determining the outflow velocity as $v_{\rm out}=2\times FWHM/2.355$. On average, the re-estimated outflow velocities are 1.3 times higher than those reported in \cite{Marasco:2023}.
At high redshift,  \cite{Concas:2022} reported only a tentative detection of outflows at $z\sim2$ in the mass range $10^8~{\rm M_{\odot}} <M_\star < 10^{9.6}~{\rm M_{\odot}}$ by stacking the spectra of 20 main-sequence galaxies, and a recent study by  \cite{Llerena:2023} reported evidence of outflow signatures in three $M_\star < 10^{9}~{\rm M_{\odot}}$ galaxies at $z\sim3$ with deep Keck/MOSFIRE observations.

The outflow velocity in our sample is about twice higher than the average outflow velocities inferred in a sample of local dwarf galaxies with similar stellar masses \citep{Marasco:2023}, suggesting that the outflow velocities increase from $z\sim0$ to $z\sim2-4$. 
Conversely, little or no evolution is found at higher redshifts. The broad component observed by \cite{Concas:2022} in the stacked spectrum of a sample of star-forming galaxies at $z\sim2$ shows a tentative outflow with a velocity of $\sim450$~\kms\ that is slightly higher than the median outflow velocity ($\sim350$~\kms) of our sample. Focussing only on the outflows at $z>4$ detected with JWST in our JADES sample, we find no evolution with redshift. A similar outflow velocity trend with cosmic time is also reported in other studies targeting massive galaxies (\mstar$\sim10^{10}~{\rm M_{\odot}}$), which showed a rapid evolution up to $z\sim2$, while the correlation is almost flat at higher redshift \citep{Sugahara:2019}.

The theoretical predictions by \cite{Nelson:2019}, who analysed outflow properties in about 20,000 galaxies with stellar mass $> 10^{7}~{\rm M}_{\odot}$ and $0.2<z<10$ in TNG50 simulations,  indicate that outflow velocities should increase with redshift, consistent with our observational trend, although the velocities at $z\sim4-8$ predicted by simulations for low-mass galaxies are $\sim200$~\kms\, which is  1.5 times lower than our median outflow velocity. In the simulations, the redshift dependence is caused by the adopted feedback prescription that connects the outflow velocity to the halo virial mass \citep{Pillepich:2018b}  in such a way that the wind velocity and the growth of the virial halo mass have the same scaling with redshift.
The statistics in our sample are still too limited to conclude whether the outflow prescription adopted in TNG50 is correct.

We also examined the relation between outflow kinematic parameters and galaxy properties, but found no correlation between the outflow velocity and star formation rate or stellar mass (see Appendix~\ref{app:outflow_prop}). The lack of a correlation is overall consistent with simulations and theoretical models, which only predict a trend in massive galaxies, where bright AGN drive fast-outflowing gas, while no trend or only a weak correlation is expected in the low-mass regime \citep{Muratov:2015, Nelson:2019}.

To verify whether outflows can represent the feedback mechanism required by theory to remove gas from galaxies and thus deplete the fuel for star formation, we compared the outflow velocity with the escape velocity from the galaxy  and dark matter halo system. 

Some observations \citep{Jones:2021, Parlanti:2023, de_Graaff:2023} and simulations \citep{Pillepich:2019, Kohandel:2019, Kohandel:2023} have indicated that low-mass ($M_{\star}<10^9~{\rm M_\odot}$) galaxies are irregular and more turbulent (rotation-to-dispersion ratio $<4$) than the rest of the population ($\sim10$), we expect that a spherical potential would match the gravitational potential of these galaxies better. We thus computed the escape velocity by considering a Hernquist potential (\citealt{Hernquist:1990a}) for the galactic component, defined by the following density profile:
\begin{equation}
    \rho(r) = \frac{M_\star+M_{\rm gas}}{2 \pi} \frac{a}{r} \frac{1}{(r+a)^3},
\end{equation}
where $a=(1+\sqrt2) r_{1/2}$, $r_{1/2}$ is the half-light (or effective) radius of the galaxy, and M$_{\rm gas}$ is the total gas mass. For the  M$_{\rm gas}$, we assumed $M_{\rm gas}/M_{\rm star}\sim1$ , which is the average gas fraction measured at $z=3-5$ \citep{Pavesi:2019, Dessauges-Zavadsky:2020}.  We used the $r_{1/2}$ determined by NIRCam imaging \citep{Rieke:2023} and reported in the JADES public catalogue\footnote{https://archive.stsci.edu/hlsp/jades}.

\begin{figure}
   \centering
   \includegraphics[width=0.5\textwidth]{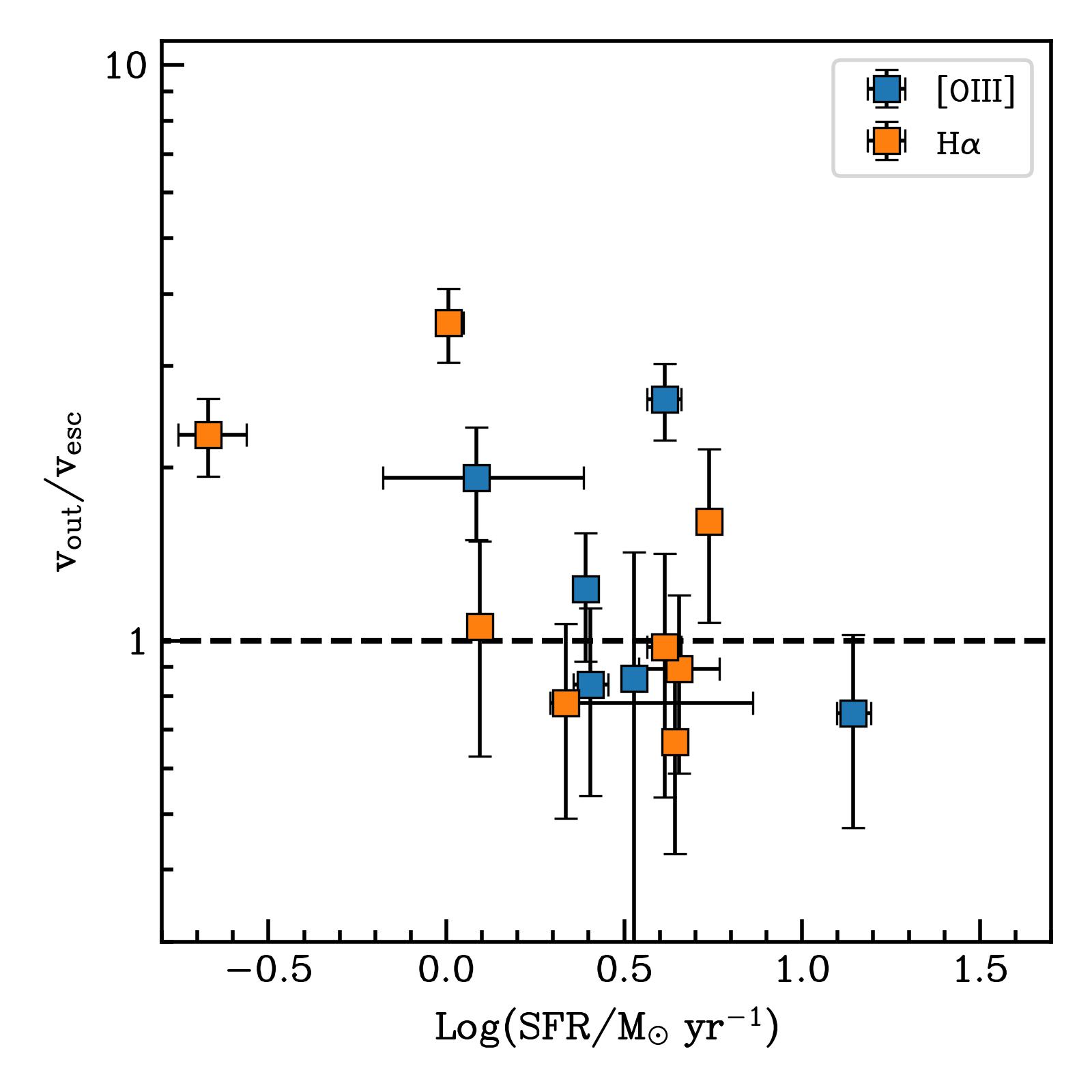}

   \caption{Ratio of the ionised outflow velocity and escape velocity vs. SFR. JADES outflows are reported with blue and orange squares. The escape velocities have been estimated by considering a Hernquist potential for the stellar and gaseous component and an NFW potential for the dark matter halo.}
    \label{fig:vout_vs_vesc}%
    \end{figure}

For the dark matter halo, we adopted a Navarro-Frenk-White (NFW) potential \citep{Navarro:1996} given by the following density profile:
\begin{equation}
    \rho(r) = \frac{\rho_\mathrm{crit}(z) \delta_\mathrm{c}}{(r/r_\mathrm{s}) (1+r/r_\mathrm{s})^2},
\end{equation}
where $\rho_\mathrm{crit}$ is the critical density of the Universe, $\delta_\mathrm{c}$ is the characteristic overdensity for the halo, and $r_\mathrm{s}$ = $r_\mathrm{200}/c(z)$ is the characteristic radius, $c(z)$ being the concentration parameter at redshift $z$. The virial dark matter halo mass ($M_{\rm DM}$) is inferred from the stellar-to-halo mass relation of \cite{Moster:2013}  at $z=0$\footnote{The uncertainties on the outflow properties and escape velocity are larger than the cosmic evolution of the stellar-to-halo mass relation, and thus, we used the local relation, which is complete in the stellar mass range covered by our sample.}, following the prescription by \cite{Posti:2019}, and the concentration $c(z)$ was determined from the $M_{\rm DM}$-$c(z)$ relation of \cite{Dutton:2014}. Finally, we determined the escape velocity for the combination of the two profiles with the python package \texttt{galpy} \citep{Bovy:2015}.
We thus found that the outflow velocities are three times higher than the escape velocity on average (Figure~\ref{fig:vout_vs_vesc}). This indicates that the outflows are able to eject gas from the galaxy and enrich the circumgalactic gas outside the virial radius and intergalactic medium on large scales. A similar result has recently been reported by \cite{Ubler:2023} in a massive $z=5.5$ AGN host galaxy, revealing an outflow of 700 \kms, which is potentially able to escape the potential well of the galaxy and enrich the intergalactic medium.
We stress that the escape velocity estimated above is on the one hand, valid for ballistic motions and thus does not take into account that the outflow might continue to be accelerated, as in the case of direct radiation pressure or of an expanding shocked bubble, while on the other hand, it is expected to be slowed down by ram pressure exerted by the ambient gas it encounters during its propagation.

\subsection{Mass-loading factor}

In the previous section, we discussed the kinematics and the potential impact of ionised outflows on low-mass galaxies at high redshift. We now determine whether the amount of ionised gas removed by outflows per unit of time is larger than the amount that is converted into stellar mass through star formation. Specifically, we estimate the mass-loading factor $\eta$, which is defined as the ratio of the mass-loss rate due to outflows and the SFR. If $\eta>1$, outflows can have a negative feedback impact on their host galaxy, effectively quenching star formation, as models predict \citep[e.g.,][]{Muratov:2015, Nelson:2019}.  

For a uniformly filled conical outflow, the mass outflow rate is defined as 
\begin{equation}
\dot{M}_{\rm out}= M_{\rm out} v_{\rm out} / r_{\rm out},
\label{eq:rate}
\end{equation} 
where $M_{\rm out}$ is the mass of the outflowing gas, and $r_{\rm out}$  is the extension of the outflow \citep[e.g.,][]{Maiolino:2012, Gonzalez-Alfonso:2017}. The mass of the gas can be estimated from the luminosity of the broad component of \ha\ \citep[e.g.,][]{Concas:2022} or \oiii\ \citep[e.g.,][]{Carniani:2015},  
\begin{equation}
M_{\rm out} =  0.8\times10^5 \left( \frac{L^{\rm corr}_{\rm H\alpha}}{\rm 10^{40}~erg~s^{-1}}\right)\left( \frac{n_{\rm out}}{\rm 100~cm^{-3}}\right)^{-1}~{\rm M_{\odot}},
\label{eq:moutha}
\end{equation} 
and 
\begin{equation}
M_{\rm out} = 0.8\times10^8 \left( \frac{L^{\rm corr}_{\rm [OIII]}}{\rm 10^{44}~erg~s^{-1}}\right)\left( \frac{Z_{\rm out}}{\rm Z_{\odot}}\right)^{-1}\left( \frac{n_{\rm out}}{\rm 500~cm^{-3}}\right)^{-1}~{\rm M_{\odot}},
\label{eq:moutoiii}
\end{equation} 

where $Z_{\rm out}$ and $n_{\rm out}$ are the metallicity and the electron density of the outflowing gas, respectively. 
 $L^{\rm corr}_{\rm H\alpha}$ and $L^{\rm corr}_{\rm [OIII]}$ are the  dust-corrected  luminosities of the $H_\alpha$ and ${\rm [OIII]}$ broad components. More specifically, we estimated the ${\rm H{\alpha}/H\beta}$ line ratio from either PRISM or R1000  data\footnote{https://archive.stsci.edu/hlsp/jades} and determined the correction factor by adopting the \cite{Calzetti:2000} attenuation law and assuming the theoretical ratios  ${\rm 
 H\alpha/H\beta=2.86}$ for case B recombination at $T= 10^4K$. We then applied the correction factors to the fluxes of the broad components.
 
 As the equations~\ref{eq:moutha} and \ref{eq:moutoiii} show, the outflow mass is sensitive to the electron density and metallicity of the outflowing gas. These two gas properties are generally assumed for most outflow studies in the local and distant Universe \citep{Forster:2019,Concas:2019,Concas:2022}. Key diagnostic emission lines needed to estimate electron density (e.g. [SII]$\lambda\lambda$6716,6731; \citealt{Forster:2019, Davies:2020}) and gas-phase metallicity ([O III]$\lambda$4363; \citealt{Cameron:2021}) are  either too faint to be detected in the observed outflows or are not covered by the G395H/F290LP NIRSpec observations for most of our targets. 
 Therefore, we assumed an electron density of $380~\rm{cm^{-3}}$, which is the typical value estimated from deep observations of $z \sim 2$ star formation-driven outflows (\citealt{Forster:2019}; also adopted by \citealt{Concas:2022})  and consistent with the electron density distribution determined by \cite{Marasco:2023} in galactic outflows of a local sample of dwarf galaxies. We then set $Z_{\rm out}$ as high as the gas-phase metallicity of the interstellar medium. The latter was recently estimated for the same JADES galaxies as  are analysed in this work by \cite{Curti:2023}  and spans a range of 
 $7.63<12+\log({\rm O/H})<7.92$. We associated a 0.3 dex uncertainty with the measurement of the ionised gas outflow mass, which takes into account the typical variations of metallicity and electron density (following \citealt{Concas:2022}). 

Determining the outflow extension $r_{\rm out}$ from the NIRSpec MSA data requires detailed modelling of the 2D spectra to quantify the impact of the background subtraction process, slit losses, and impact of the bar shadow on the surface brightness emission. In addition, a detailed kinematic analysis is necessary to distinguish the galactic disk emission from the outflow emission pixel by pixel. In this work, which is meant as a first exploration of outflows in low-mass galaxies at high redshift, we therefore used the same assumption as was adopted in other studies, which assumed the ionised outflows to be as extended as the galaxy, that is, $r_{\rm out}=r_{\rm gal}$. This assumption is supported by the typical sizes of ionised gas outflows determined with high spatial resolution observations of local and low-$z$ star-forming galaxies \citep{Newman:2012, Forster:2014}. We thus adopted the half-light radius of the galaxies determined by NIRCam imaging \citep{Rieke:2023} to calculate the mass-loss rate of the ionised outflows.

\begin{figure}
   \centering
   \includegraphics[width=0.5\textwidth]{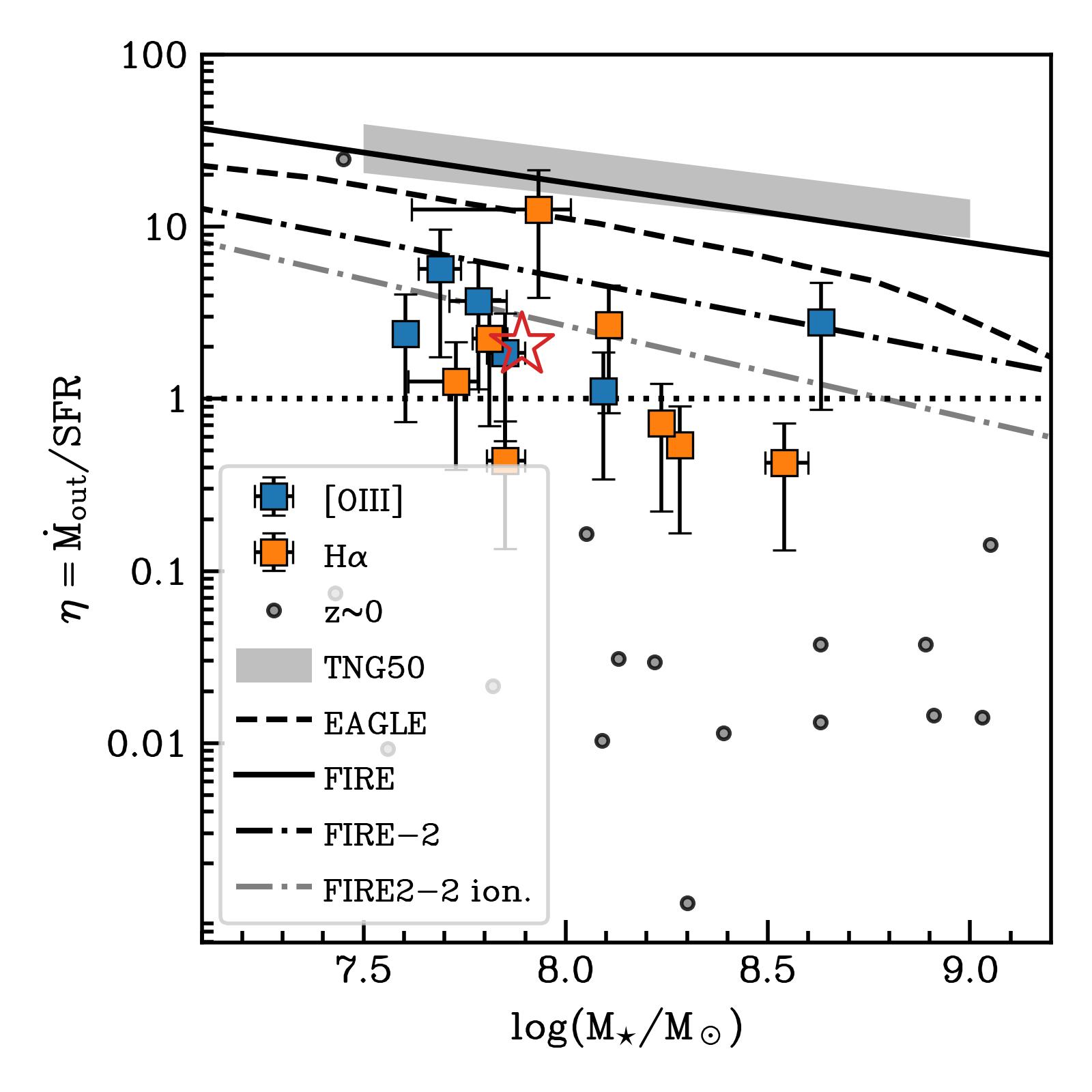}

   \caption{Stellar mass dependence of the outflow mass-loading factor (i.e. the ratio of the mass-loss rate and SFR). Estimates based on \oiii\ and \ha\ emission lines are reported as blue and orange squares, respectively. The median outflow mass-loading factor is illustrated as a red star. The grey circles show the mass-loading factors from a sample of dwarf galaxies at $z\sim0$ \citep{Marasco:2023}.
   The shaded grey region shows the prediction from the TNG50 simulations at $z\sim5$ \citep{Nelson:2019}, and the solid and dashed black lines are the trend  at $z\sim5$  from EAGLE \citep{Mitchell:2020} and the redshift-independent relation from FIRE simulations \citep{Muratov:2015}, respectively. The dot-dashed black and grey lines show the prediction from the FIRE-2 simulation for all outflow phases combined and for  the ionised outflow alone, respectively \citep{Pandya:2021}.}
              \label{fig:mass_loading_factor}%
\end{figure}
    
Based on the inferred estimates of outflow mass, velocity, and radius, we determined the mass outflow rate by using Equation~\ref{eq:rate}
 and, from this, the mass-loading factor for each outflow (Table~\ref{tab:tab_outflow}). Our sample has a median mass-loading factor of $\eta=2.0^{+1.6}_{-1.5}$, which agrees well with models that predict that feedback from supernovae is the main outflow driver and is required to regulate star formation and metal enrichment in galaxies \citep[e.g.][]{Finlator:2008, Dave:2011, Heckman:2015}. 

Figure~\ref{fig:mass_loading_factor} presents the dependence of the
outflow mass-loading factor on stellar mass for the JADES sample. The figure shows that $\eta$ is inversely proportional to stellar mass. It reaches values up to $\eta=2-12$ at $M_\star< 10^{8}~{\rm M_\odot}$ and decreases to $\sim0.3$ at $M_\star\sim10^{8.5}~{\rm M_\odot}$. 
Theory predicts that outflows driven by a momentum- or energy-conserving shocked expanding bubble powered by star formation activity are expected to have a power-law dependence of the mass loading factor on stellar mass with slopes in the range $-0.33$ to $-0.6$ \citep{Dutton:2012, Muratov:2015, Nelson:2019, Pandya:2021}. This is different from the steeper trend we find (Figure~\ref{fig:mass_loading_factor}), whose slope is $-0.9_{-0.1}^{+0.5}$. 
%

In addition, the inferred mass-loading factor is five times lower than the predictions from cosmological simulations on average, which require $\eta= 10$  in the mass range of our sample to reproduce the properties of galaxies in the local Universe \citep{Muratov:2015, Nelson:2019, Mitchell:2020, Pandya:2021}.  This discrepancy may arise from the fact that optical nebular lines only map the warm ($T\sim10^4$ K) ionised gas, while part of the gas may be in a cold molecular and neutral atomic phase  ($T\sim10^2$ K) or even in a hot ($T\gtrsim10^6-10^8$ K) ionised phase \citep{Cicone:2014, Carniani:2015, Nelson:2019, Fluetsch:2019, Herrera-Camus:2019, Fluetsch:2021}. We indeed note that our mass-loading factors are more consistent with the theoretical expectations when we only consider the warm phase of the outflowing gas predicted by the FIRE-2 simulation (dot-dashed grey line in Fig. ~\ref{fig:mass_loading_factor}).

In addition to the multiphase nature of  outflows, we need to consider that the fraction of ionised gas at a given distance from the galaxy depends on the number of ionising photons intercepting the gas clouds at that distance, and consequently, the ionised gas fraction is determined by the covering factor of the gas clouds (i.e., the fraction of sky covered by photoionised gas clouds as seen from the emitter) and by the shading of the flux due to the gas clouds at smaller distances. Therefore, the tracer we used (the warm ionised gas) may not map the bulk of outflowing gas at large distances. 
Finally, the mass-outflow rate estimated based on the nebular lines could be underestimated as the photo-centroid of the light emission of both \ha\ and \oiii\ could be at smaller distances, and the assumption for the outflow radius used in Equation~\ref{eq:rate} may be overestimated. 
Deep NIRSpec observations in integral-field spectroscopic (IFS) mode will be fundamental to spatially resolve ionised outflows in the early Universe and determine their properties as a function of distance from the galactic centre.

In conclusion, the mass-loading factor estimated from the rest-frame optical nebular lines is at least as large as the star formation rate estimates for most of the galaxies in our sample. We note that the median mass-loading factor is 100 times higher than what is observed in the local Universe by \cite{Marasco:2023}, who reported a median mass-loading factor of 0.02. This indicates that the star formation activity in the early Universe can be strongly affected by  galactic outflows, which, according to theoretical studies, can be driven by either supernovae or radiation pressure from stars, or by both \citep{Heckman:1990, Murray:2005, Dutton:2012,  Muratov:2015, Hopkins:2016, Hopkins:2018,Nelson:2019, Ziparo:2023}. 
However, we cannot exclude that faint accreting super-massive black holes are hidden at the centres of the galaxies in our sample, and their radiation boosts the outflow energetics, as predicted by recent theoretical works  \citep{Dashyan:2018, Koudmani:2019, Mezcua:2023}.

\section{Conclusions}
\label{sec:conclusions}

We investigated the outflow incidence and properties  of 52 low-mass ($M_\star<10^{10}$~M$_{\odot}$) star-forming galaxies at $z > 3$ by exploiting the JWST/NIRSpec R2700 observations of the JADES Deep/HST program. We inferred the outflowing gas through the analysis of the rest-frame optical lines, specifically, \ha\ or \oiii, or both, which can be detected with NIRSpec up to $z\sim9$.  Our main findings are listed below.

   \begin{enumerate}
      \item We find evidence for ionised outflows in $25\%-40\%$ of the 52 galaxies of the sample, where the incidence rate depends on the S/N. This incidence rate may be caused by the geometry of the outflowing gas. When the outflows have a biconical morphology with an opening angle of $\sim45$~deg, we expect an incidence rate of $\sim30\%$, which is consistent with our observations.
      
      \item The incidence of outflows increases with stellar mass and SFR, suggesting that outflows are more frequent in more massive and starbursting galaxies. 

      \item The inferred outflow velocities are 1.5 times higher than the median velocity observed in local dwarf star-forming galaxies. This indicates that outflows are more powerful in the distant Universe than they are at low redshift.

      \item For some galaxies, the velocity of the outflowing gas is higher than the escape velocity from the gravitational potential of the host galaxy and its dark matter halo. This means that the expelled outflows are able to enrich both the circumgalactic and the intergalactic media. 

        \item The inferred mass-loading factor of the outflows (mass-loss rate versus star formation rate in the host) spans values between $\eta=0.4-12$, with a median of $2$, which is about 100 times higher than the mass-loading factors observed in local dwarf galaxies. This indicates that the impact of ejective feedback mechanisms in the early galaxies is significant and might explain the presence of non-star-forming galaxies in the first billion years of the Universe \citep{Looser:2023}.
        
        \item We find that the mass-loading factor decreases with increasing stellar mass with a slope of $-1.3$.  Although an anti-correlation is consistent with the predictions from theoretical studies, the mass-loading factor is a factor of 3 smaller than that predicted by simulations for galaxies  at the same masses and redshifts.

   \end{enumerate}

This study provides the first census of ionised outflows in the galaxy population at $z \sim 3-9$ with stellar masses $10^7~{\rm M_{\odot}}<M_\star<10^9$~M$_{\odot}$. Our results suggest that ionised outflows play a crucial role in the evolution of galaxies and in the metal enrichment of the circum- and intergalactic media. The JWST enables us to detect and study outflows on a wide range of stellar masses and SFRs in the high-redshift Universe and provide new insights on galaxy formation and evolution. Future deep and wide NIRSpec observations will be crucial to expand the current sample of galaxies showing outflows and understand their impact on galaxies.

\begin{table*}
\caption{Properties of the detected outflows and of their host galaxies.}
\label{tab:tab_outflow}
\begin{tabular}{l c c c c c c c c}      
\hline\hline  
{\small ID} &  {\small $z$} & {\small Log($M_\star$/M$_\odot$)} & {\small Log(SFR/\sfr)} & {\small Log($M_{\rm out}$/M$_\odot$)} & {\small $r_{\rm out}$} & {\small $v_{\rm out}$} & {\small Log($\dot M_{\rm out}/{\rm M_\odot~yr^{-1}})$} & {\small Log($\eta$)} \\
   &        &  &  & &  [kpc] & [\kms] &   & \\

(1) & (2) & (3) & (4) & (5) & (6) & (7) & (8) & (9)\\

\hline  
 \multicolumn{8}{c}{\oiii}   \\
\hline
5457 & 4.861 &$7.69\pm0.05$ & $0.09\pm0.19$ & $6.46\pm0.02$ & 0.21 & $500\pm55$ & $0.84\pm0.30$ & $0.8\pm0.3$ \\
9422 & 5.935 &$7.60\pm0.02$ & $0.53\pm0.00$ & $7.07\pm0.01$ & 0.30 & $234\pm77$ & $0.91\pm0.30$ & $0.4\pm0.3$ \\
18090 & 4.773 &$7.85\pm0.05$ & $0.61\pm0.02$ & $6.84\pm0.02$ & 0.64 & $701\pm53$ & $0.88\pm0.30$ & $0.3\pm0.3$ \\
18846 & 6.334 &  $8.09\pm0.02$ & $0.39\pm0.01$ & $6.56\pm0.01$ & 0.52 & $401\pm50$ & $0.43\pm0.30$ & $0.0\pm0.3$ \\
10013609 & 6.928 &$7.78\pm0.07$ & $0.41\pm0.02$ & $7.12\pm0.01$ & 0.33 & $259\pm46$ & $0.97\pm0.30$ & $0.6\pm0.3$ \\
10013704 & 5.919 & $8.63\pm0.02$ & $1.14\pm0.02$ & $8.26\pm0.01$ & 0.31 & $289\pm53$ & $2.22\pm0.30$ & $1.1\pm0.3$ \\
\hline  
 \multicolumn{8}{c}{\ha}   \\
\hline
3184 & 3.467 &$8.54\pm0.05$ & $0.65\pm0.06$ & $6.74\pm0.01$ & 0.71 & $267\pm46$ & $0.29\pm0.30$ & $-0.4\pm0.3$ \\
4270 & 4.022 & $8.11\pm0.02$ & $0.74\pm0.01$ & $7.17\pm0.02$ & 0.44 & $444\pm74$ & $1.17\pm0.30$ & $0.4\pm0.3$ \\
6246 &5.560 &  $7.73\pm0.12$ & $0.14\pm0.05$ & $6.00\pm0.02$ & 0.29 & $497\pm44$ & $0.24\pm0.30$ & $0.1\pm0.3$ \\
7762 & 4.148 & $8.28\pm0.01$ & $0.34\pm0.02$ & $6.54\pm0.01$ & 0.61 & $229\pm42$ & $0.07\pm0.30$ & $-0.3\pm0.3$ \\
7892 & 4.228 & $7.81\pm0.04$ & $0.09\pm0.01$ & $6.03\pm0.01$ & 0.10 & $275\pm56$ & $0.45\pm0.30$ & $0.4\pm0.3$ \\
17260 & 4.885 & $7.93\pm0.31$ & $-0.67\pm0.04$ & $5.85\pm0.02$ & 0.17 & $648\pm50$ & $0.43\pm0.30$ & $1.1\pm0.3$ \\
18090 & 4.773 & $7.85\pm0.05$ & $0.61\pm0.02$ & $6.67\pm0.01$ & 0.64 & $261\pm59$ & $0.25\pm0.30$ & $-0.4\pm0.3$ \\
10016186 & 3.927 & $8.24\pm0.01$ & $0.01\pm0.02$ & $6.51\pm0.02$ & 4.06 & $911\pm67$ & $-0.13\pm0.30$ & $-0.1\pm0.3$ \\
\hline
\end{tabular}
\\{\bf Note:} (1) NIRSpec ID of the target; (2) stellar mass; (3) star formation rate over the last 10 Myr; (4) mass outflow rate assuming $n_e= 380~{\rm cm^{-3}}$ (as in \citealt{Forster:2019} and \citealt{Concas:2022} for $z\sim2$ outflows); (5) outflow radius; (6) outflow velocity; (7) mass outflow rate; (8) mass-loading factor.
\end{table*}

\begin{acknowledgements}

 We thank the anonymous referee for constructive comments and
suggestions; Marasco A. for providing data and useful information about his line-fitting analysis of local dwarf-galaxies; Yi Xu for useful discussion and comments on the escape velocity estimates
SC and GV acknowledge support by European Union’s HE ERC Starting Grant No. 101040227 - WINGS.
ECL acknowledges support of an STFC Webb Fellowship (ST/W001438/1).
RM, JS, JW, and LS acknowledge support by the Science and Technology Facilities Council (STFC) and by the ERC through Advanced Grant 695671 "QUENCH". RM also acknowledges funding from a research professorship from the Royal Society.
JW also thanks the support by the Fondation MERAC.
AJB, AJC, JC acknowledge funding from the "FirstGalaxies" Advanced Grant from the European Research Council (ERC) under the European Union’s Horizon 2020 research and innovation programme (Grant agreement No. 789056).
RS acknowledges support from a STFC Ernest Rutherford Fellowship (ST/S004831/1).
SA, MP and BRP acknowledge support from Grant PID2021-127718NB-I00 funded by the Spanish Ministry of Science and Innovation/State Agency of Research (MICIN/AEI/ 10.13039/501100011033). 
MP acknowledges support from  the Programa Atracci\'on de Talento de la Comunidad de Madrid via grant 2018-T2/TIC-11715.
H{\"U} gratefully acknowledges support by the Isaac Newton Trust and by the Kavli Foundation through a Newton-Kavli Junior Fellowship.
DJE is supported as a Simons Investigator and by JWST/NIRCam contract to the University of Arizona, NAS5-02015.
BER acknowledges support from the NIRCam Science Team contract to the University of Arizona, NAS5-02015. 
The research of CCW is supported by NOIRLab, which is managed by the Association of Universities for Research in Astronomy (AURA) under a cooperative agreement with the National Science Foundation.
Funding for this research was provided by the Johns Hopkins University, Institute for Data Intensive Engineering and Science (IDIES).
This research is supported in part by the Australian Research Council Centre of Excellence for All Sky Astrophysics in 3 Dimensions (ASTRO 3D), through project number CE170100013.
The Cosmic Dawn Center (DAWN) is funded by the Danish National Research Foundation under grant no.140.

\end{acknowledgements}

%
%
\bibliographystyle{aa}
\bibliography{bibliography}

\begin{appendix} 

\section{Spectroscopic analysis and best-fit results}
\label{app:best_fitting_results}

\begin{figure}
   \centering
   \includegraphics[width=0.5\textwidth]{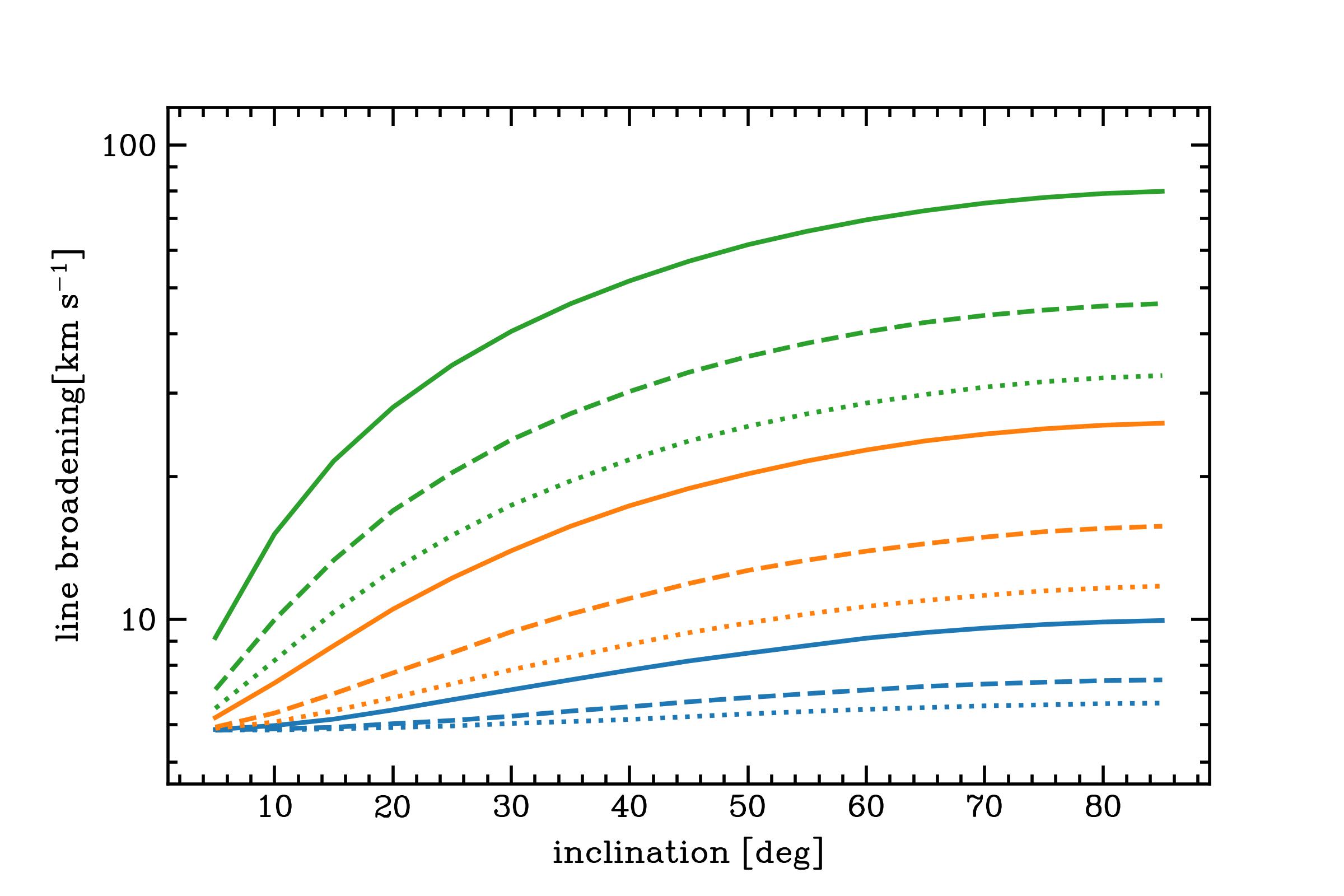}
   \caption{Line broadening (i.e. velocity dispersion) of the integrated mock 1D spectra for different models of rotating disks as a function of disk inclination with respect to the line of sight. The blue, orange, and green lines show the line broadening for galaxies with masses of 10$^7$~M$_\odot$, 10$^8$~M$_\odot$, and 10$^9$~M$_\odot$, respectively. Disk models with a half-light radius of 0.1~kpc, 0.3~kpc, and 0.7~kpc are reported with continuous, dashed, and dotted lines, respectively.}
              \label{fig:sigma_vs_inc}%
    \end{figure}

 As reported in Sec.~\ref{sec:outflow_identification}, line broadening in the spatially integrated 1D spectra can be caused by the rotational motions of a galactic disk. To test this possibility, we determined the maximum line broadening associated with a rotating disk for a typical low-mass galaxy at $z>3$. We generated mock 1D spectra by using the code \texttt{KinMS} \citep{Davis:2013} for different dynamical masses, galaxy scale radii, and disk inclinations with respect to the line of sight. We assumed that the disk surface mass density distribution is exponential and used equation (6) in \cite{Parlanti:2023}  to determine the circular velocity as a function of radius \cite{Binney:2008}. Figure~\ref{fig:sigma_vs_inc} shows the emission line broadening due to circular motions as a function of the galactic disk
inclination, dynamical mass, and scale radius. In the mass range of our sample, the line broadening (i.e. velocity dispersion) does not reach values higher than 80~\kms.

After we set the constraints of the priors for the single- and double-Gaussian components, we performed the fitting of the data for each galaxy and determined the best-fit results and Bayesian evidence for each model.
In Figure~\ref{fig:spectra}, we present the spectral fitting of the targets showing evidence for an additional broad component based on the criteria defined in Sec.~\ref{sec:outflow_identification}. For each galaxy, we report the best-fit results for the two- and single-Gaussian models.
The best-fit results for the modelling with two Gaussian profiles are reported in Tables~\ref{tab:table1} and ~\ref{tab:table1b}.
We note that the widths of the broad component are larger than the maximum broadening of 80~\kms\ expected from rotation. We thus conclude that the additional broad component does not trace the rotation motion in the galaxy, but maps the fast-emitting gas likely associated with outflows.

   \begin{figure}
   \centering
   \includegraphics[width=0.5\textwidth]{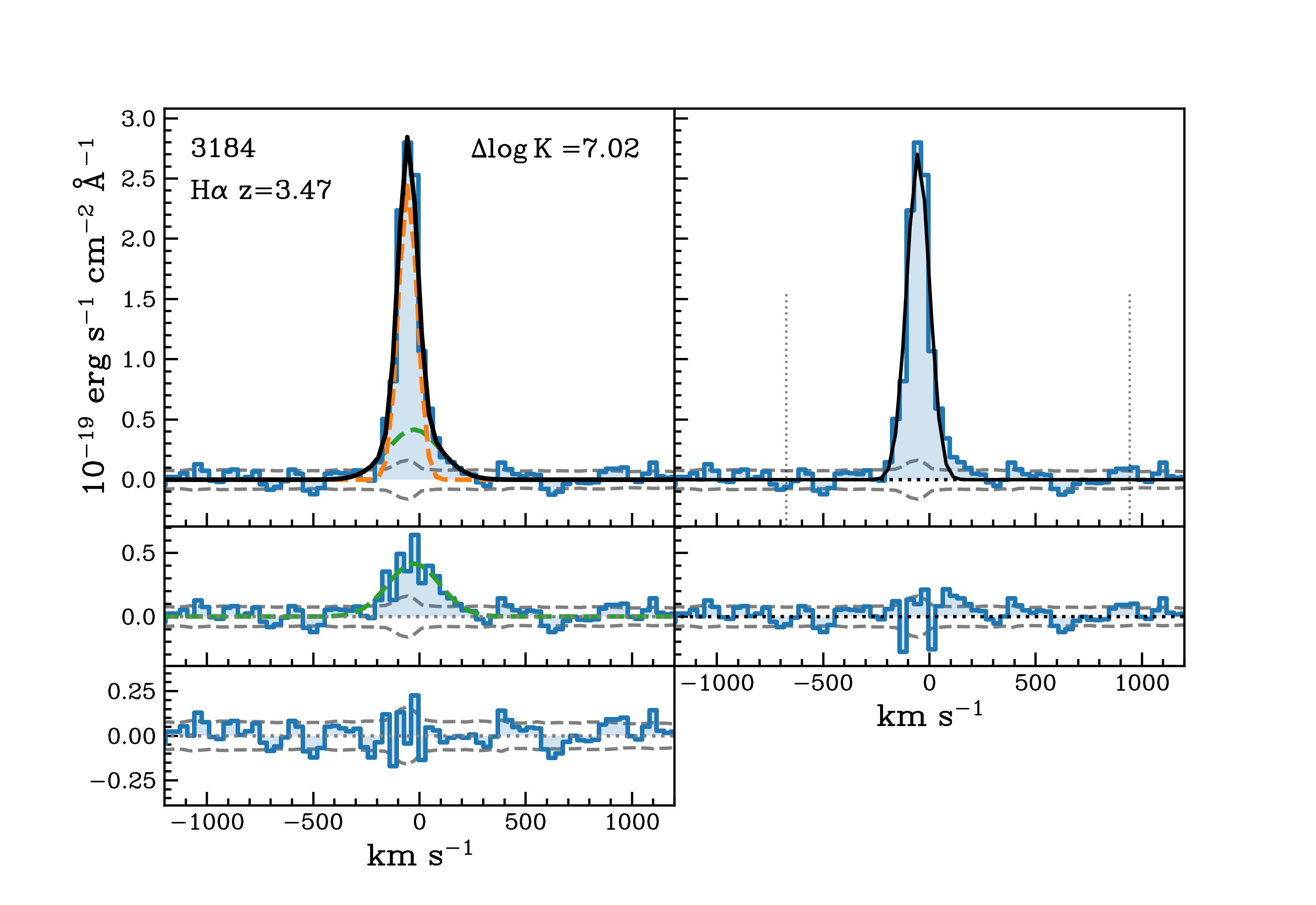}
   \includegraphics[width=0.5\textwidth]{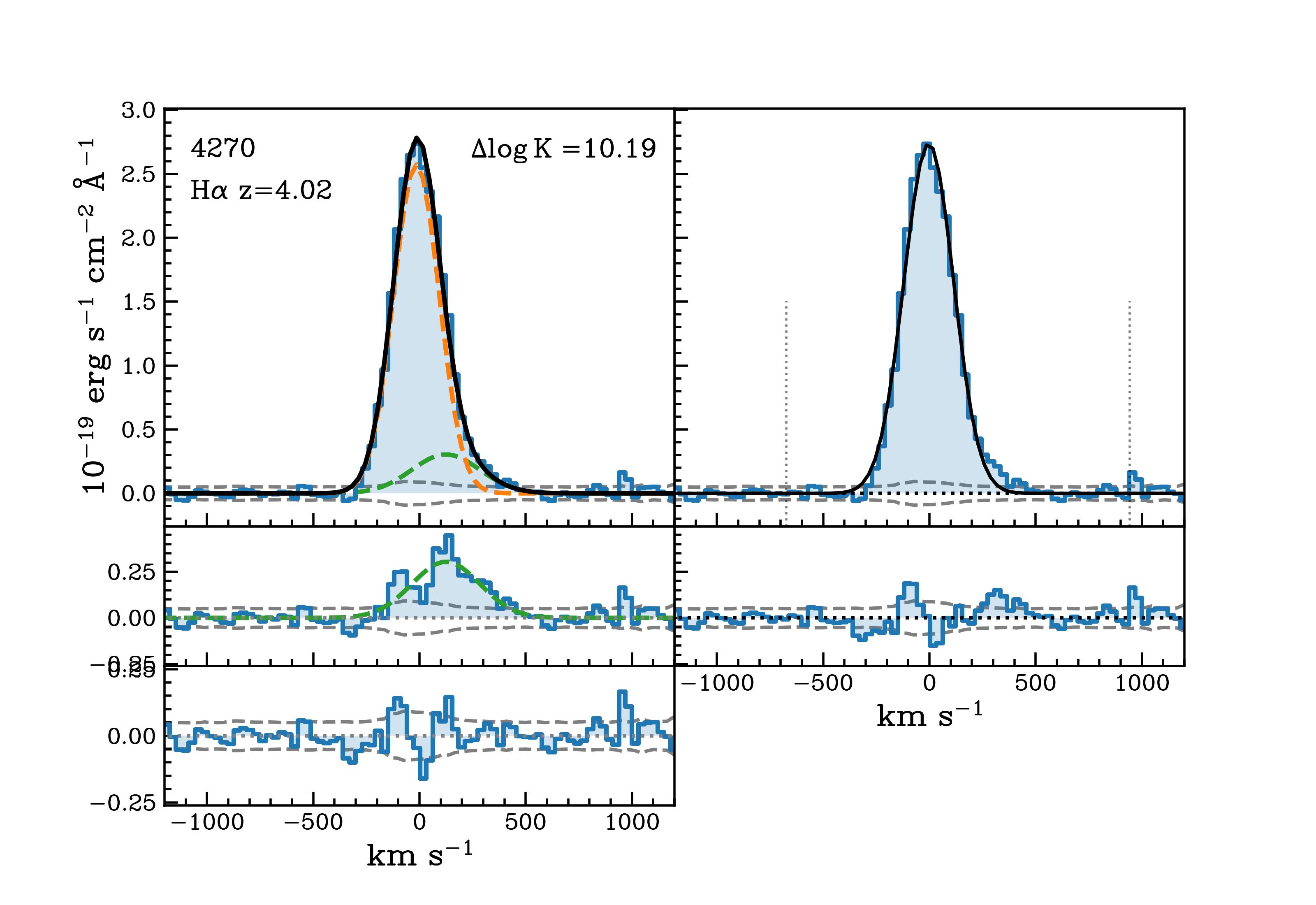}
   \includegraphics[width=0.5\textwidth]{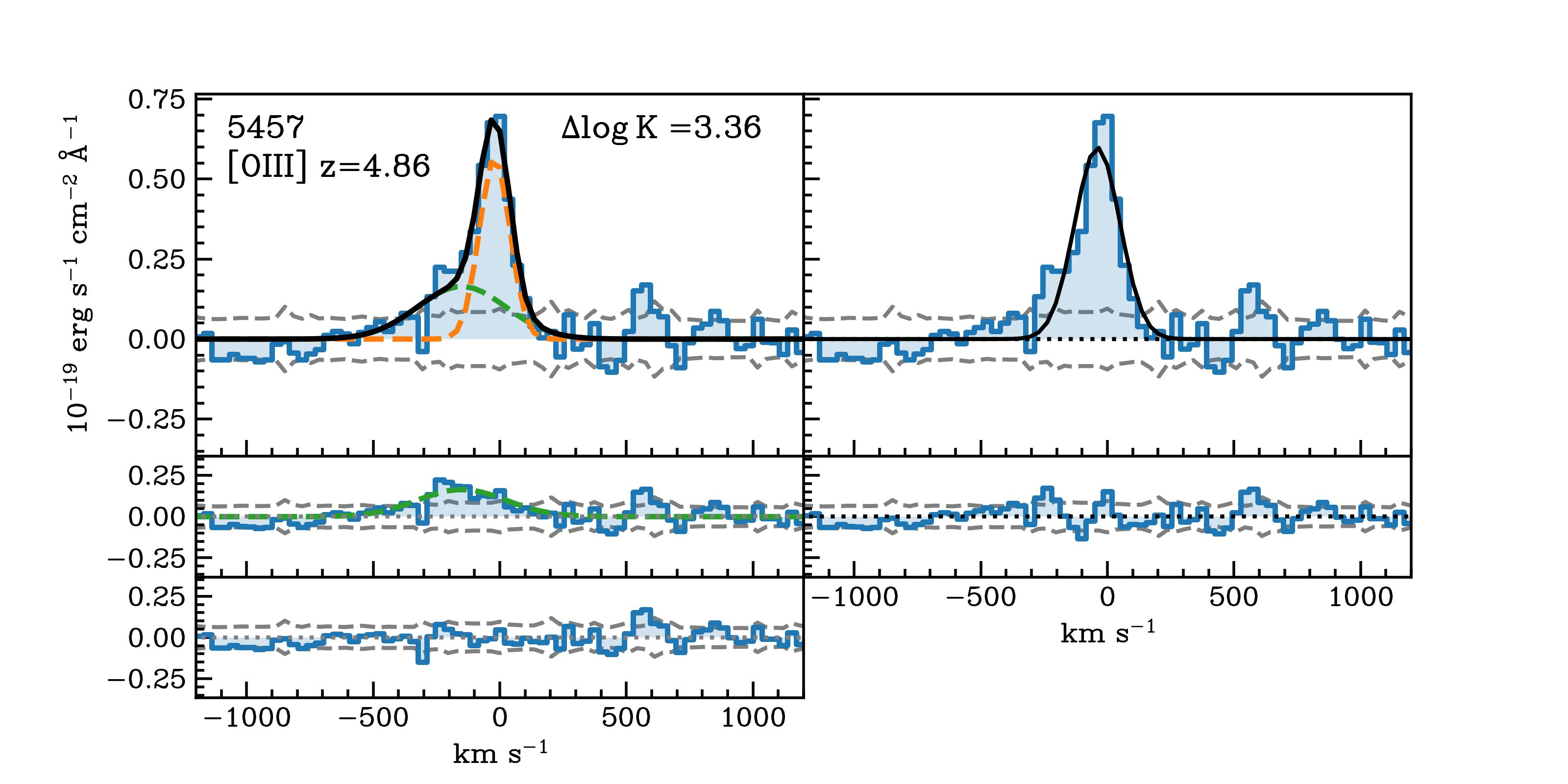}
   \caption{Rest-frame optical emission line profiles of the targets showing outflow features. For each target, the top left and right panels show the two- and single-Gaussian best-fit models in black, respectively. The text in the left panel reports the ID of the target, its redshift, the name of the rest-frame optical emission line, and  the difference between the logarithmic Bayesian evidence ($\Delta\log K$) of the two models. The second panel from the top reports the residuals from the subtraction of the core component (the orange line in the top left panel and the black line in the top right panel). The third panel from the top illustrates the residuals from the double-Gaussian component fitting. The $\pm1\sigma$ rms of the spectra is indicated by dashed grey lines.}
              \label{fig:spectra}%
    \end{figure}

\begin{figure}
   \centering
      \includegraphics[width=0.5\textwidth]{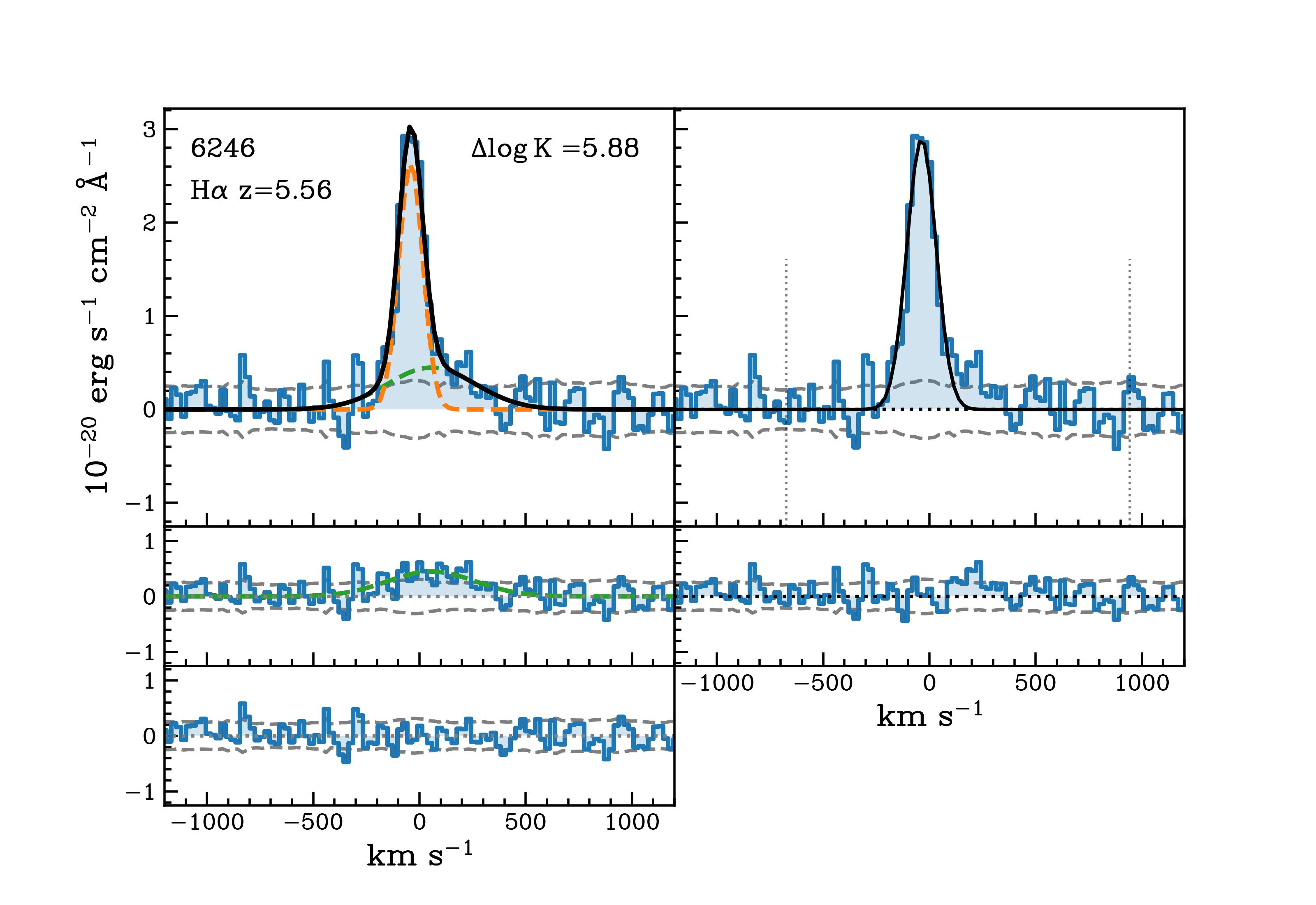}
      \includegraphics[width=0.5\textwidth]{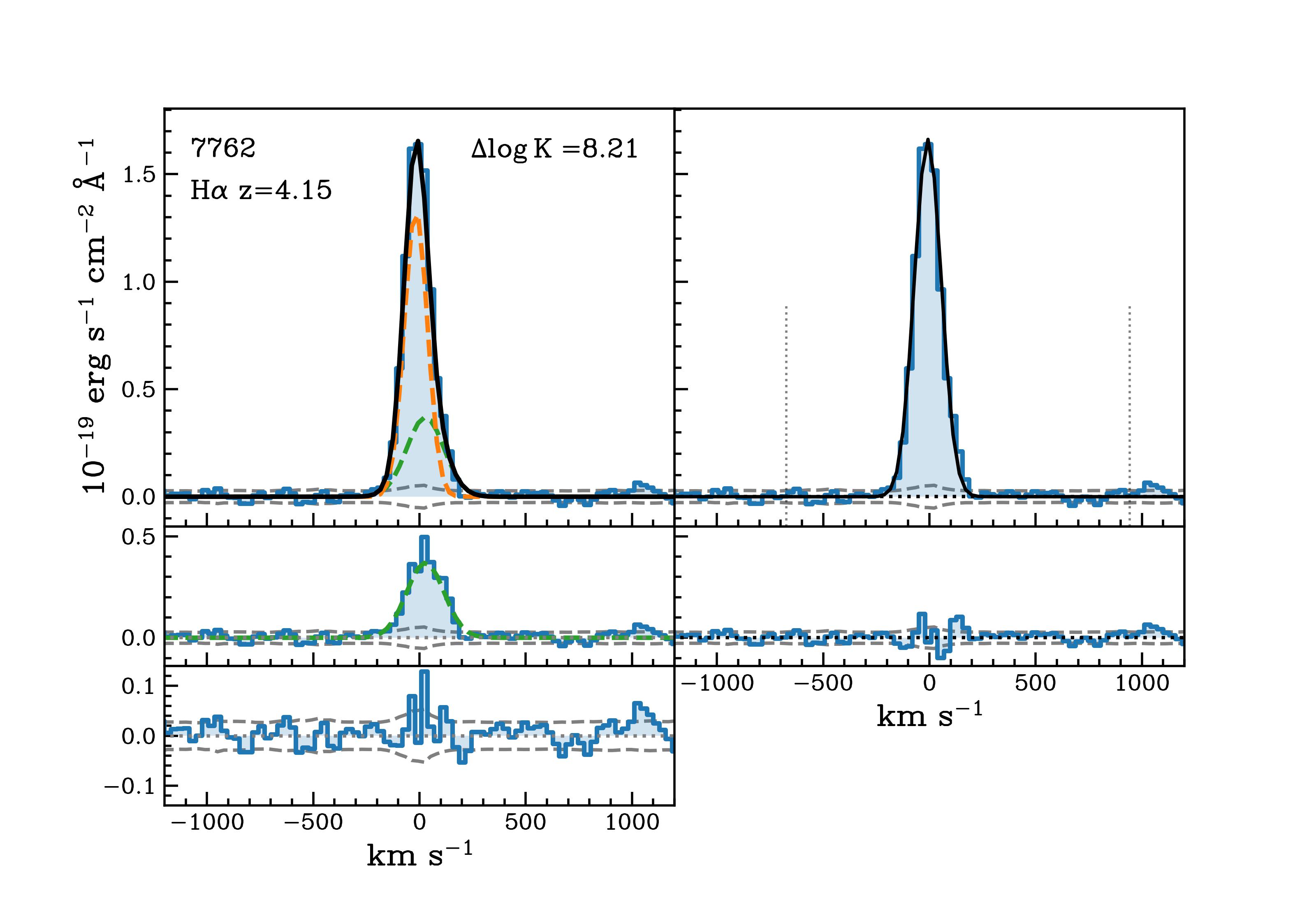}
    \includegraphics[width=0.5\textwidth]{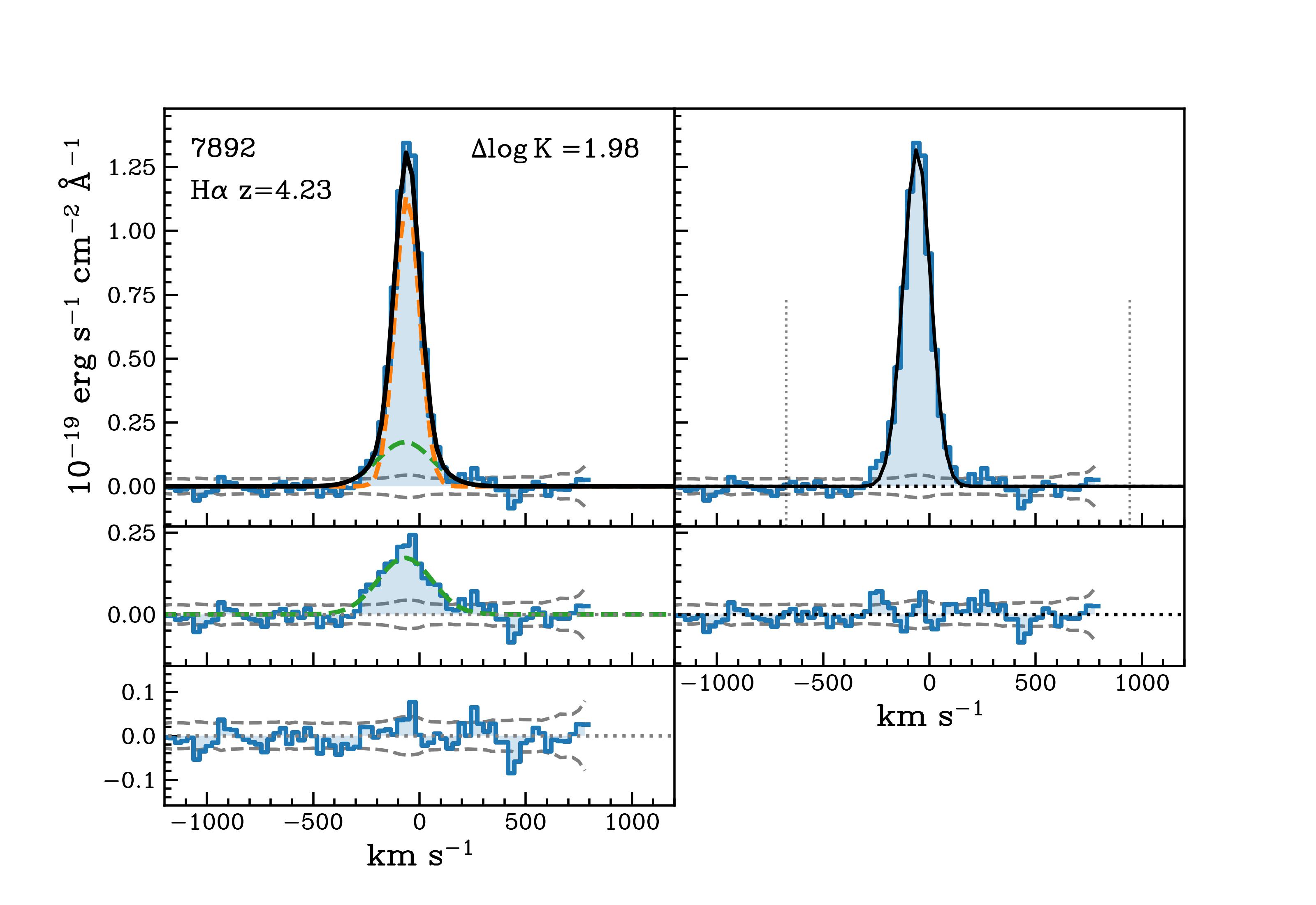}
   \includegraphics[width=0.5\textwidth]{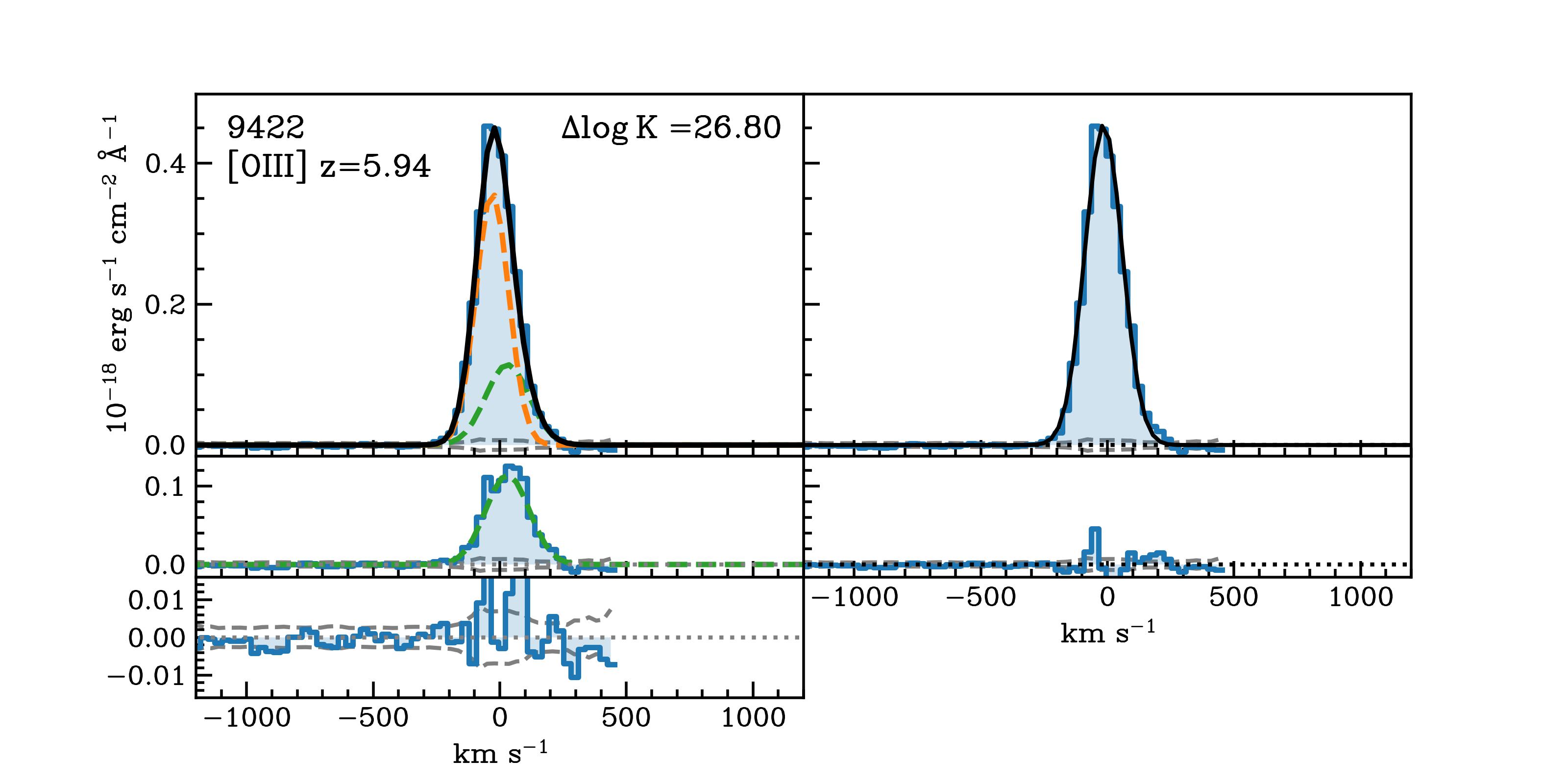}

   \caption{continued.}
    \end{figure}
    \addtocounter{figure}{-1}

\begin{figure}
   \centering
      \includegraphics[width=0.49\textwidth]{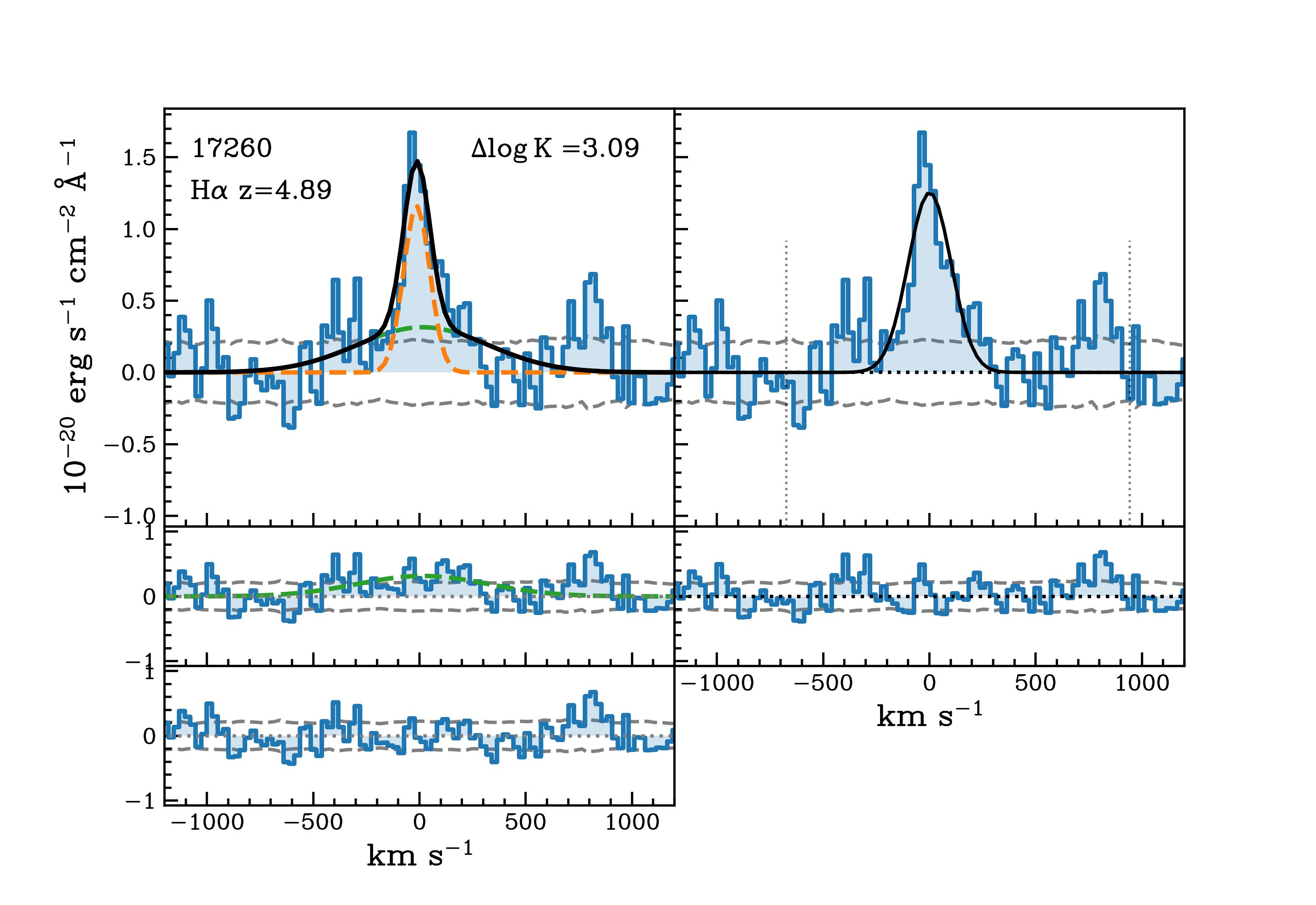}      \includegraphics[width=0.49\textwidth]{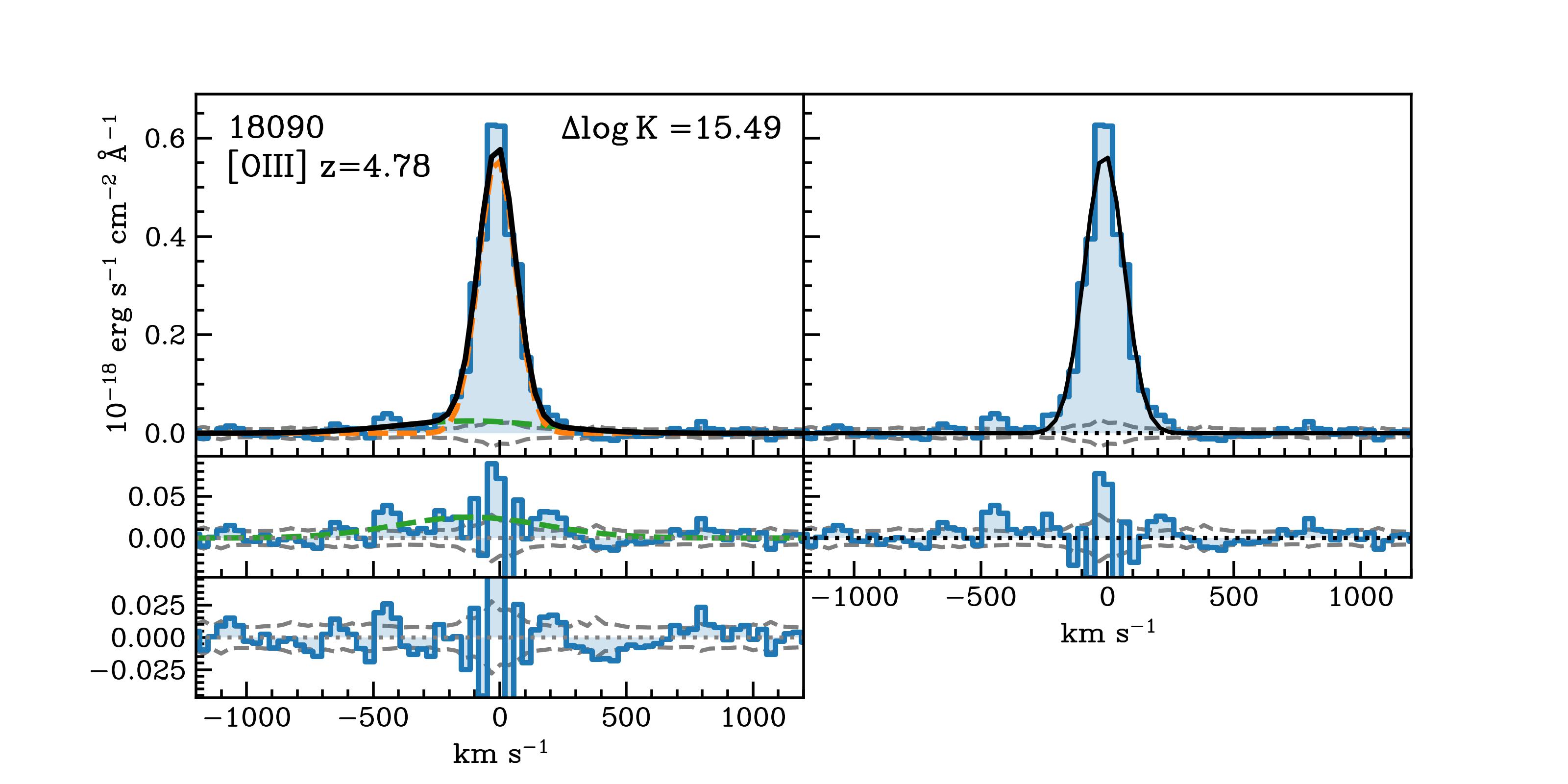}
       \includegraphics[width=0.49\textwidth]{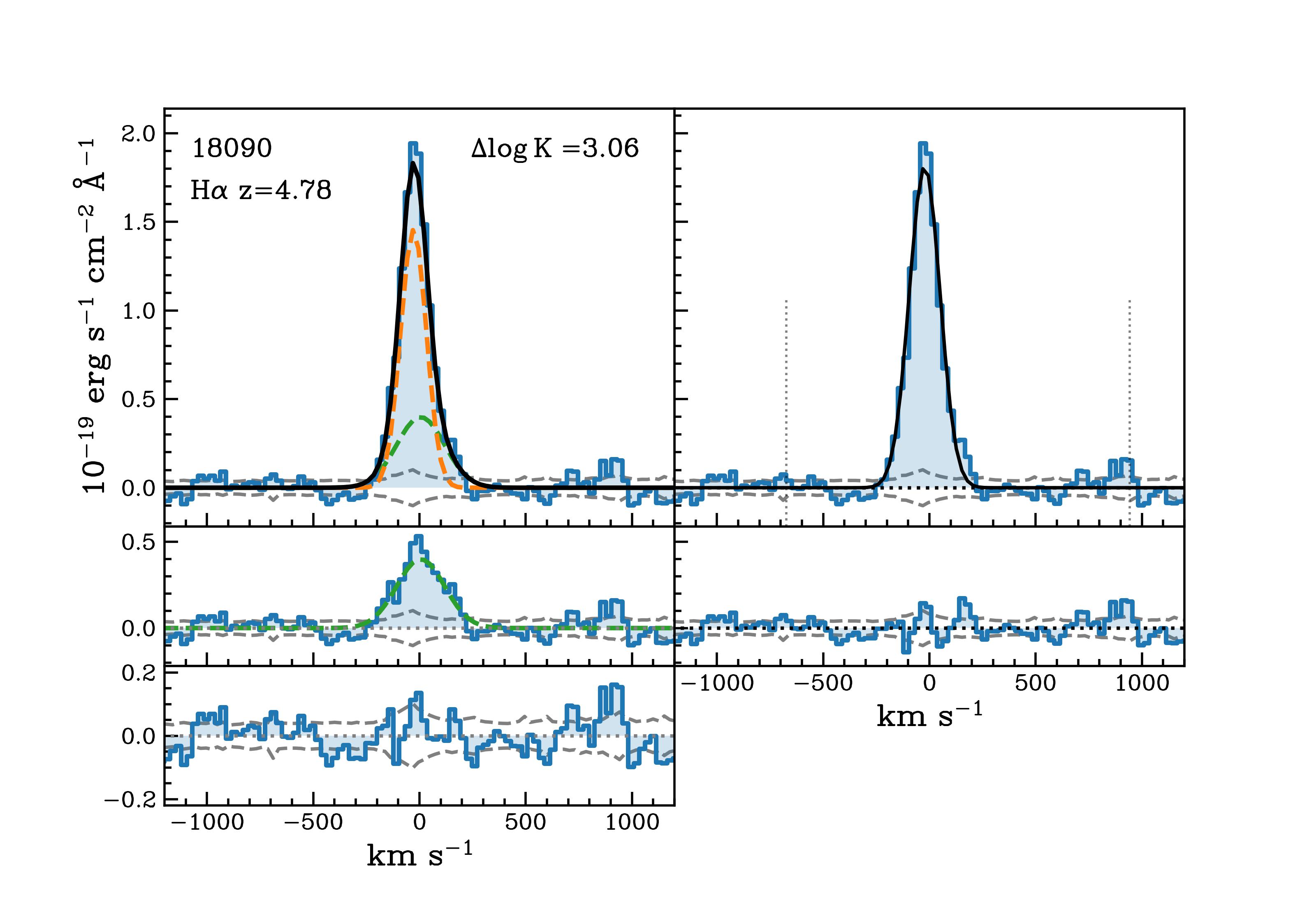}

     \includegraphics[width=0.49\textwidth]{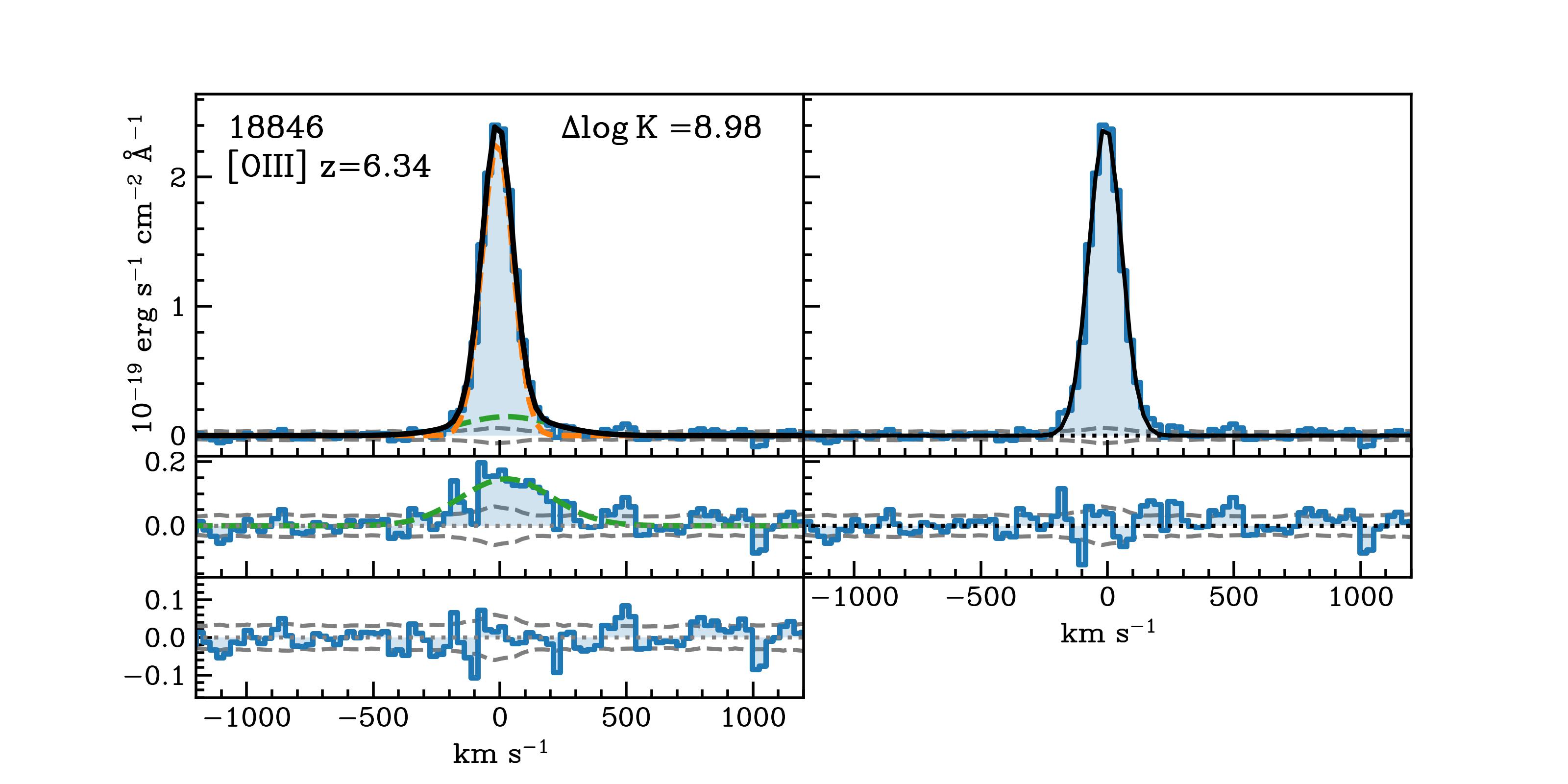}

   \caption{continued.}
    \end{figure}
    \addtocounter{figure}{-1}

\begin{figure}
   \centering
   \includegraphics[width=0.5\textwidth]{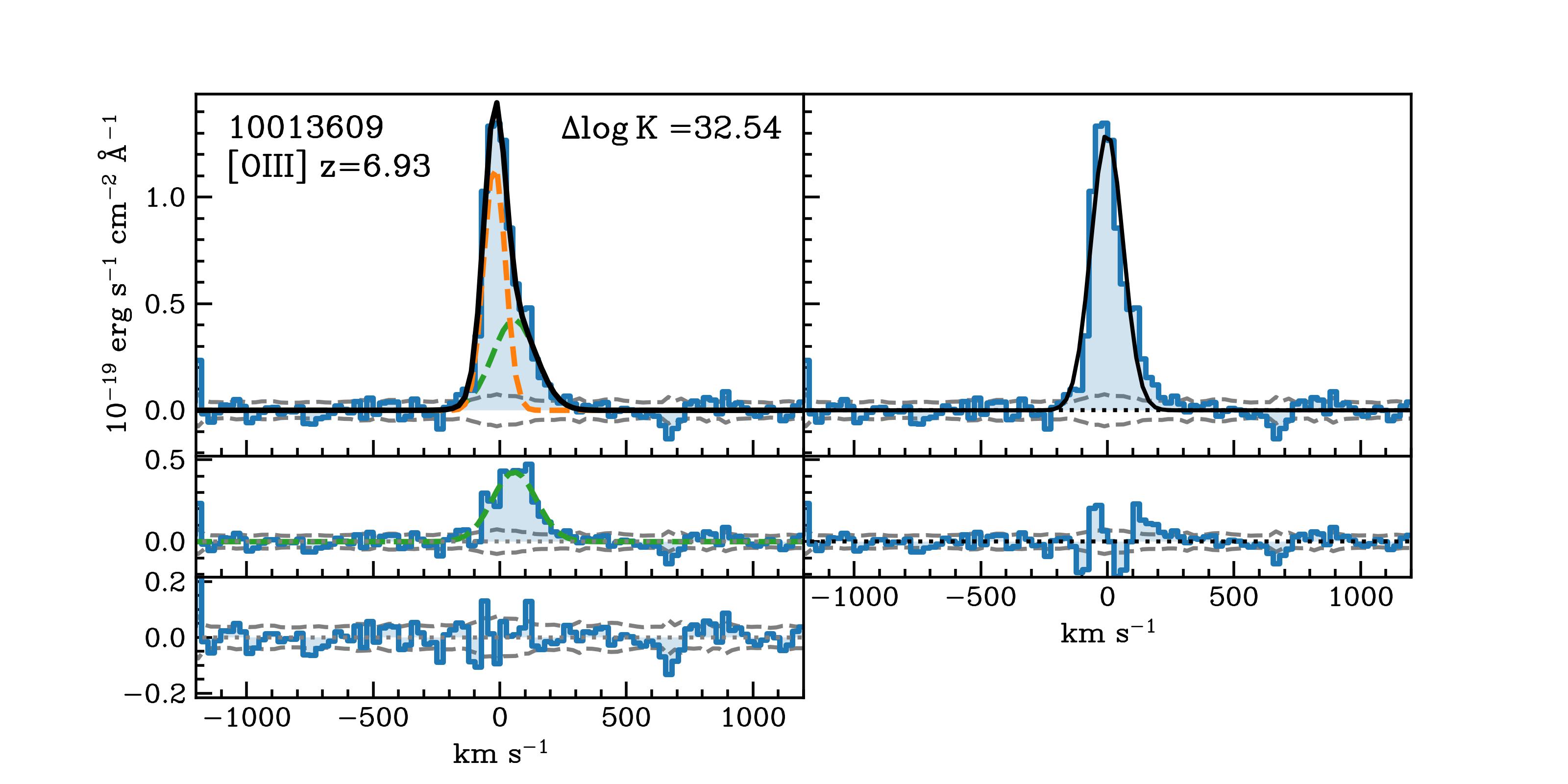}
  \includegraphics[width=0.5\textwidth]{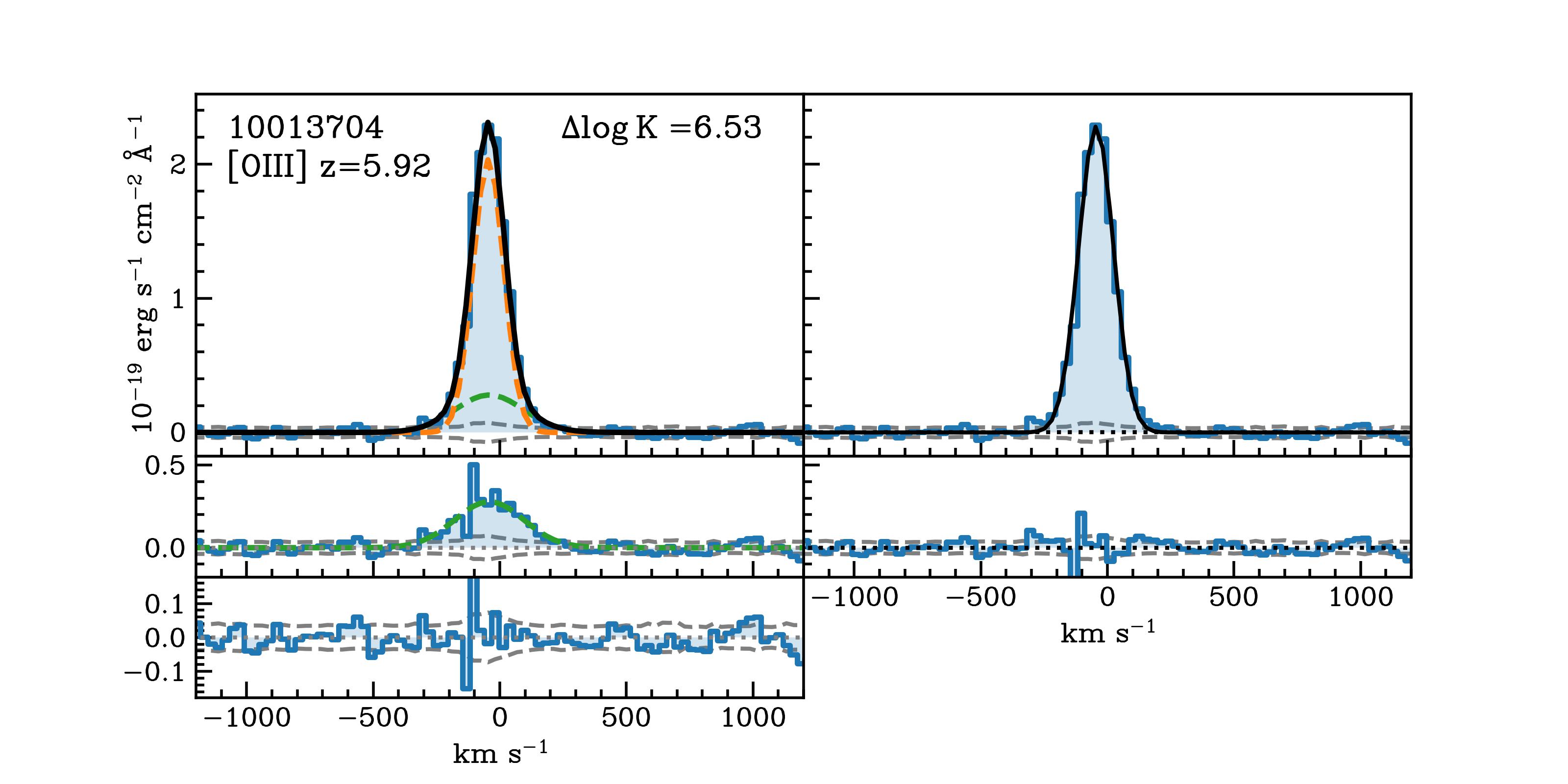}

   \includegraphics[width=0.5\textwidth]{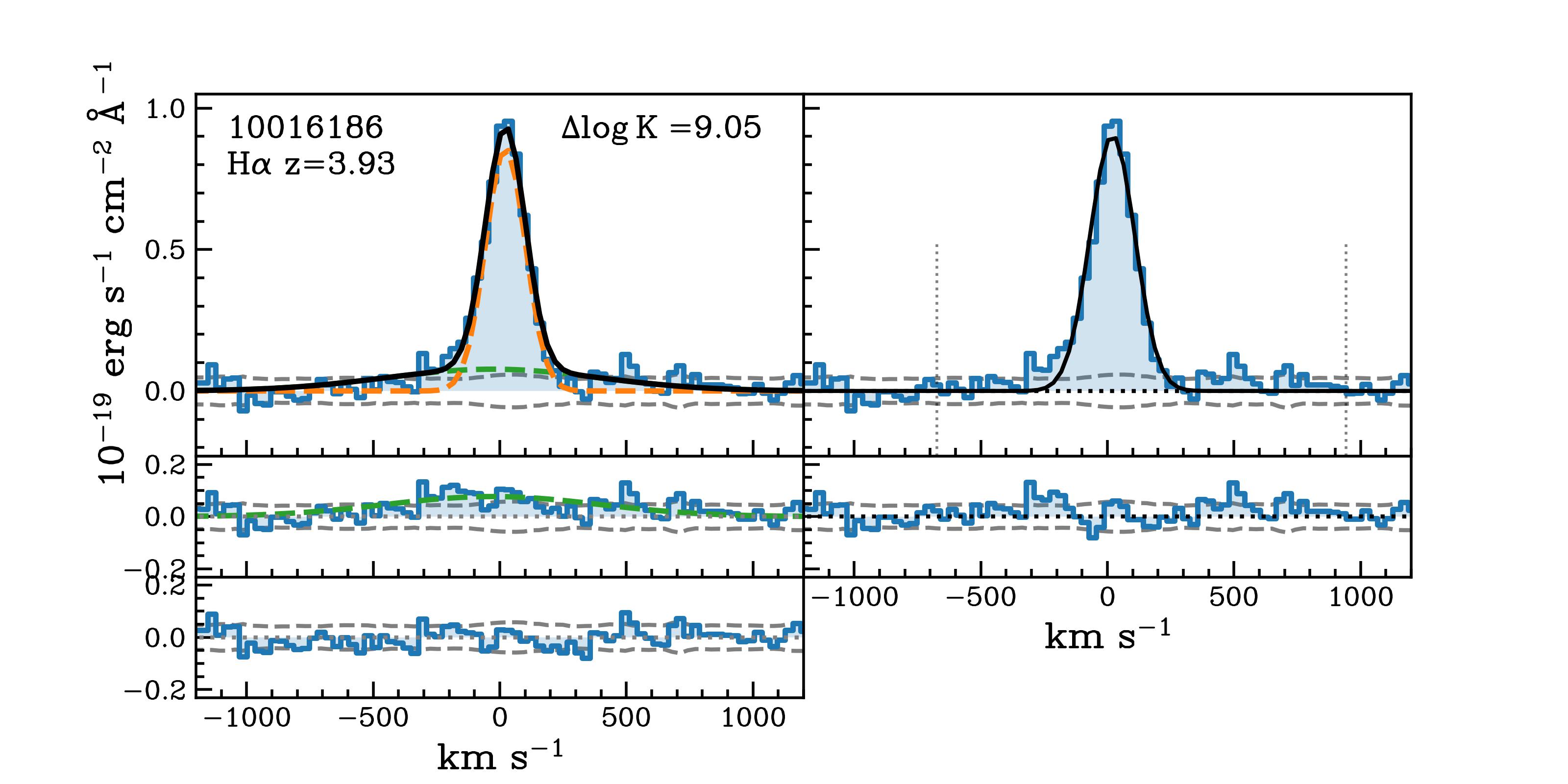}     
   \caption{continued.}
    \end{figure}
    \addtocounter{figure}{-1}

\begin{landscape}
\centering
\begin{table}
\caption{\ha\ best-fit results. }             
\label{tab:table1}      
\begin{tabular}{l c c c c c c c c c c} 
\hline\hline       
ID & $A_{\rm narrow}$ & $v_{\rm narrow}$ & $\sigma_{\rm narrow}$ & 
$\delta_{\rm A}$ & $v_{\rm broad}$ & $\delta_{\rm sigma}$ & S/N$_{\rm broad}$ & $\chi^2_{\rm single}$ & $\chi^2_{\rm double}$  & $\Delta \log K$ \\ 
   & [${\rm 10^{-19}~erg~s^{-1}~cm^{-2}~\AA^{-1}}$] & [${\rm km~s^{-1}}$] & [${\rm km~s^{-1}}$] 
   & & [${\rm km~s^{-1}}$] & & & & & \\ 
 (1)  & (2) &  (3) & (4) & (5)  & (6) & (7) &(8)  & (9) & (10) &  (11)\\ 

\hline                    
3184 & $2.40\pm0.04$ & $-54.9\pm43.1$ & $44.3\pm0.1$ & $0.1865\pm0.0014$ & $-29.3\pm14.2$ & $2.72\pm0.05$ & 10.8 & 1.41 & 0.85 & 7.29 \\
4270 & $2.51\pm0.10$ & $-16.1\pm71.2$ & $108.7\pm5.0$ & $0.1419\pm0.0009$ & $107.4\pm12.9$ & $1.47\pm0.00$ & 5.1 & 1.43 & 0.75 & 10.33 \\
4297 & $0.62\pm0.01$ & $-30.7\pm28.6$ & $43.4\pm0.0$ & $0.1089\pm0.0005$ & $-126.1\pm3.2$ & $2.90\pm0.01$ & 1.2 & 0.74 & 0.66 & -1.89 \\
4404 & $1.09\pm0.03$ & $-12.2\pm55.7$ & $63.2\pm1.5$ & $0.0137\pm0.0000$ & $45.2\pm3.1$ & $4.31\pm0.01$ & 0.4 & 0.90 & 0.92 & -3.67 \\
5329 & $0.45\pm0.02$ & $-3.5\pm49.3$ & $65.8\pm2.2$ & $0.1101\pm0.0029$ & $197.6\pm14.0$ & $8.05\pm0.07$ & 1.9 & 0.98 & 0.81 & 1.36 \\
5759 & $0.59\pm0.01$ & $34.5\pm36.0$ & $55.0\pm0.9$ & $0.0640\pm0.0005$ & $-344.9\pm1.3$ & $4.72\pm0.00$ & 0.9 & 1.09 & 0.99 & -1.26 \\
6002 & $0.35\pm0.01$ & $-15.9\pm49.2$ & $58.7\pm1.5$ & $0.1221\pm0.0014$ & $114.8\pm13.3$ & $1.98\pm0.00$ & 1.7 & 1.44 & 1.31 & 0.61 \\
6246 & $0.26\pm0.01$ & $-41.1\pm40.5$ & $57.4\pm1.2$ & $0.1659\pm0.0026$ & $51.8\pm15.1$ & $3.52\pm0.03$ & 2.7 & 0.95 & 0.69 & 3.92 \\
6384 & $0.38\pm0.01$ & $61.1\pm31.6$ & $48.1\pm0.1$ & $0.0872\pm0.0004$ & $75.7\pm1.9$ & $4.12\pm0.01$ & 0.7 & 1.30 & 1.29 & -1.89 \\
7507 & $1.11\pm0.03$ & $31.3\pm65.4$ & $69.4\pm2.6$ & $0.0371\pm0.0001$ & $33.9\pm9.7$ & $3.05\pm0.01$ & 1.1 & 1.00 & 0.98 & -3.55 \\
7762 & $1.38\pm0.03$ & $-16.6\pm40.7$ & $55.9\pm1.7$ & $0.2383\pm0.0005$ & $29.3\pm9.9$ & $1.64\pm0.00$ & 14.5 & 1.24 & 0.64 & 7.43 \\
7809 & $1.10\pm0.05$ & $-37.2\pm57.5$ & $117.6\pm5.4$ & $0.0675\pm0.0006$ & $-70.6\pm2.8$ & $4.49\pm0.00$ & 1.0 & 0.86 & 0.70 & -1.46 \\
7892 & $1.14\pm0.03$ & $-57.5\pm53.9$ & $58.2\pm1.7$ & $0.1608\pm0.0010$ & $-68.5\pm10.5$ & $2.26\pm0.04$ & 8.2 & 0.96 & 0.58 & 3.16 \\
7938 & $1.14\pm0.02$ & $-26.4\pm52.1$ & $60.5\pm1.7$ & $0.0595\pm0.0012$ & $59.8\pm7.0$ & $3.09\pm0.02$ & 2.5 & 1.55 & 1.36 & 1.49 \\
8083 & $3.66\pm0.09$ & $-48.5\pm55.8$ & $62.5\pm2.4$ & $0.0690\pm0.0013$ & $-78.0\pm14.4$ & $3.88\pm0.09$ & 10.5 & 4.70 & 1.44 & 85.81 \\
8113 & $0.84\pm0.02$ & $45.6\pm53.0$ & $52.9\pm1.0$ & $0.0348\pm0.0010$ & $-345.9\pm0.8$ & $7.89\pm0.04$ & 1.7 & 0.97 & 0.82 & 0.82 \\
9422 & $2.05\pm0.03$ & $-11.1\pm59.0$ & $65.9\pm1.9$ & $0.0290\pm0.0003$ & $130.5\pm10.7$ & $2.01\pm0.00$ & 1.6 & 1.32 & 1.14 & -0.38 \\
9743 & $1.15\pm0.03$ & $-29.2\pm31.7$ & $56.4\pm1.4$ & $0.2357\pm0.0026$ & $-419.4\pm7.0$ & $1.45\pm0.00$ & 1.9 & 1.51 & 1.18 & 3.56 \\
16745 & $0.51\pm0.02$ & $-22.9\pm53.2$ & $69.4\pm3.0$ & $0.0639\pm0.0025$ & $204.2\pm12.8$ & $6.91\pm0.08$ & 2.0 & 1.17 & 0.86 & 6.38 \\
17072 & $0.18\pm0.00$ & $-41.4\pm32.4$ & $51.6\pm0.3$ & $0.1003\pm0.0004$ & $-120.0\pm9.0$ & $3.49\pm0.01$ & 1.2 & 1.01 & 0.95 & -1.48 \\
17260 & $0.12\pm0.01$ & $-12.7\pm47.9$ & $63.3\pm0.3$ & $0.2794\pm0.0049$ & $14.8\pm12.4$ & $4.90\pm0.06$ & 3.6 & 1.30 & 1.09 & 2.72 \\
17777 & $0.38\pm0.01$ & $-39.4\pm46.4$ & $59.8\pm1.6$ & $0.0968\pm0.0020$ & $302.2\pm15.8$ & $5.47\pm0.09$ & 2.5 & 1.18 & 0.82 & 4.37 \\
18028 & $0.38\pm0.01$ & $-9.9\pm38.2$ & $41.6\pm0.0$ & $0.0800\pm0.0007$ & $27.9\pm5.7$ & $4.81\pm0.02$ & 1.6 & 0.76 & 0.74 & -1.01 \\
18090 & $1.45\pm0.04$ & $-28.3\pm57.1$ & $61.3\pm2.6$ & $0.2971\pm0.0005$ & $2.5\pm9.9$ & $1.87\pm0.01$ & 12.8 & 1.22 & 0.90 & 2.90 \\
18970 & $4.20\pm0.11$ & $-10.9\pm46.2$ & $62.7\pm2.3$ & $0.2429\pm0.0006$ & $-58.3\pm14.1$ & $1.79\pm0.01$ & 13.3 & 2.11 & 1.81 & 1.75 \\
19519 & $2.11\pm0.03$ & $-62.1\pm54.2$ & $60.8\pm1.5$ & $0.0272\pm0.0000$ & $-13.6\pm4.9$ & $2.77\pm0.00$ & 2.2 & 1.08 & 1.07 & -4.34 \\
10000626 & $0.26\pm0.01$ & $23.0\pm43.4$ & $54.2\pm0.8$ & $0.0573\pm0.0003$ & $-40.5\pm9.7$ & $3.95\pm0.01$ & 0.9 & 0.58 & 0.56 & -1.87 \\
10005113 & $0.45\pm0.01$ & $13.4\pm36.5$ & $44.1\pm0.2$ & $0.0523\pm0.0003$ & $55.8\pm6.5$ & $3.77\pm0.01$ & 1.1 & 0.46 & 0.44 & -1.99 \\
10009506 & $2.66\pm0.06$ & $22.6\pm55.1$ & $61.2\pm1.6$ & $0.0948\pm0.0004$ & $-10.2\pm8.3$ & $2.11\pm0.02$ & 5.8 & 0.81 & 0.62 & -0.71 \\
10013545 & $2.15\pm0.03$ & $-13.0\pm35.6$ & $64.5\pm1.6$ & $0.0694\pm0.0001$ & $23.9\pm2.5$ & $2.81\pm0.00$ & 4.1 & 0.88 & 0.70 & -3.00 \\
10013704 & $0.91\pm0.03$ & $-43.5\pm45.4$ & $65.8\pm2.4$ & $0.1384\pm0.0046$ & $221.8\pm19.3$ & $4.29\pm0.14$ & 4.2 & 3.02 & 1.16 & 60.09 \\
10015338 & $0.97\pm0.02$ & $-24.9\pm52.1$ & $70.3\pm2.0$ & $0.1132\pm0.0012$ & $-175.3\pm15.7$ & $2.67\pm0.01$ & 1.6 & 1.06 & 0.95 & -0.50 \\
10016186 & $0.86\pm0.04$ & $-38.4\pm54.2$ & $80.2\pm3.3$ & $0.0897\pm0.0032$ & $-89.5\pm13.4$ & $5.36\pm0.06$ & 3.1 & 1.21 & 0.71 & 9.05 \\
10016374 & $0.66\pm0.02$ & $-14.6\pm45.1$ & $63.3\pm2.0$ & $0.1215\pm0.0002$ & $-68.3\pm10.5$ & $1.97\pm0.00$ & 2.7 & 0.80 & 0.71 & -2.53 \\
10035295 & $2.24\pm0.04$ & $19.0\pm36.3$ & $58.3\pm1.4$ & $0.1381\pm0.0002$ & $37.1\pm9.8$ & $2.11\pm0.00$ & 8.2 & 1.01 & 0.94 & -2.50 \\
\hline                  
\end{tabular}
\\{\bf Note:} (1) NIRSpec ID of the target; (2) amplitude of the narrow component; (3,4) velocity centroid and dispersion of the narrow Gaussian component with respect to the systemic redshift of the galaxy reported in \cite{Bunker:2023}; (5) broad-to-narrow Gaussian amplitude ratio; (6) velocity centroid of the broad Gaussian component with respect to the systemic redshift of the galaxy; (7)  broad-to-narrow Gaussian velocity dispersion ratio; (8) signal-to-noise ratio of the integrated flux of the broad component; (9,10) reduced chi-squared for the single- and double-Gaussian models, respectively; (11) log-difference between the two Bayesian evidences of the single and double-Gaussian models.\\\end{table}
\end{landscape}

\begin{landscape}
\centering
\begin{table}
\caption{\oiii\ best-fit results}             
\label{tab:table1b}      
\begin{tabular}{l c c c c c c c c c c}      
\hline\hline       
ID & $A_{\rm narrow}$ & $v_{\rm narrow}$ & $\sigma_{\rm narrow}$ & 
$\delta_{\rm A}$ & $v_{\rm broad}$ & $\delta_{\rm sigma}$ & S/N$_{\rm broad}$ & $\chi^2_{\rm single}$ & $\chi^2_{\rm double}$  & $\Delta \log K$ \\ 
   & [${\rm 10^{-19}~erg~s^{-1}~cm^{-2}~\AA^{-1}}$] & [${\rm km~s^{-1}}$] & [${\rm km~s^{-1}}$] 
   & & [${\rm km~s^{-1}}$] & & & & & \\ 
 (1)  & (2) &  (3) & (4) & (5)  & (6) & (7) &(8)  & (9) & (10) &  (11)\\ 
\hline
3968 & $0.37\pm0.01$ & $51.2\pm54.5$ & $64.8\pm1.7$ & $0.0498\pm0.0004$ & $-160.8\pm2.8$ & $5.67\pm0.01$ & 0.8 & 0.45 & 0.40 & -1.74 \\
4297 & $1.49\pm0.02$ & $-30.9\pm49.3$ & $57.7\pm1.0$ & $0.0178\pm0.0003$ & $-304.1\pm1.0$ & $5.55\pm0.02$ & 1.2 & 0.94 & 0.85 & -1.79 \\
4404 & $2.33\pm0.05$ & $-5.8\pm61.4$ & $64.8\pm2.2$ & $0.0250\pm0.0003$ & $-31.6\pm7.8$ & $4.18\pm0.01$ & 1.8 & 0.55 & 0.45 & -1.43 \\
5457 & $0.56\pm0.02$ & $-20.5\pm52.7$ & $60.8\pm0.8$ & $0.3017\pm0.0032$ & $-153.1\pm15.8$ & $3.02\pm0.03$ & 4.1 & 1.04 & 0.66 & 3.89 \\
5759 & $1.01\pm0.02$ & $20.1\pm39.7$ & $66.7\pm2.0$ & $0.0266\pm0.0001$ & $30.6\pm2.9$ & $5.10\pm0.01$ & 0.5 & 0.45 & 0.45 & -2.54 \\
6002 & $0.79\pm0.02$ & $-15.1\pm36.9$ & $55.4\pm1.2$ & $0.1098\pm0.0002$ & $23.8\pm11.9$ & $2.06\pm0.00$ & 4.9 & 0.79 & 0.66 & -2.17 \\
6246 & $0.55\pm0.01$ & $-44.6\pm50.0$ & $63.0\pm1.3$ & $0.1140\pm0.0002$ & $-2.9\pm6.2$ & $2.08\pm0.01$ & 3.2 & 1.08 & 1.01 & -2.47 \\
7938 & $2.57\pm0.06$ & $-17.8\pm39.5$ & $57.2\pm1.4$ & $0.2580\pm0.0006$ & $-35.0\pm9.0$ & $1.99\pm0.02$ & 15.1 & 0.76 & 0.44 & 0.15 \\
8013 & $0.50\pm0.01$ & $-11.0\pm49.8$ & $52.0\pm0.8$ & $0.0518\pm0.0003$ & $83.2\pm3.2$ & $4.31\pm0.01$ & 0.9 & 0.76 & 0.73 & -2.27 \\
8113 & $2.21\pm0.04$ & $34.9\pm54.7$ & $57.1\pm1.3$ & $0.0215\pm0.0001$ & $-17.7\pm6.6$ & $3.22\pm0.01$ & 1.3 & 0.89 & 0.89 & -3.71 \\
9422 & $3.57\pm0.12$ & $-30.1\pm75.5$ & $66.0\pm2.7$ & $0.3328\pm0.0004$ & $29.8\pm12.0$ & $1.32\pm0.00$ & 34.8 & 3.32 & 1.65 & 26.88 \\
9452 & $1.66\pm0.07$ & $-18.3\pm49.1$ & $93.8\pm3.2$ & $0.0382\pm0.0001$ & $36.7\pm6.2$ & $3.72\pm0.01$ & 0.4 & 0.85 & 0.86 & -2.89 \\
9743 & $1.92\pm0.02$ & $-32.2\pm35.7$ & $56.1\pm0.7$ & $0.0508\pm0.0001$ & $-72.4\pm5.1$ & $3.17\pm0.00$ & 1.3 & 0.48 & 0.45 & -2.71 \\
16625 & $1.23\pm0.02$ & $-0.6\pm54.5$ & $48.6\pm0.7$ & $0.0191\pm0.0001$ & $-77.2\pm3.5$ & $4.61\pm0.02$ & 1.5 & 0.72 & 0.70 & -3.15 \\
16745 & $1.39\pm0.03$ & $-27.1\pm46.7$ & $74.4\pm2.5$ & $0.0132\pm0.0002$ & $156.1\pm9.3$ & $6.04\pm0.02$ & 1.1 & 0.83 & 0.77 & -2.42 \\
17566 & $0.90\pm0.03$ & $-18.7\pm58.8$ & $81.1\pm3.3$ & $0.0974\pm0.0012$ & $50.9\pm12.9$ & $2.95\pm0.01$ & 2.2 & 1.98 & 1.80 & 1.74 \\
18090 & $5.62\pm0.12$ & $-10.9\pm44.7$ & $73.7\pm3.0$ & $0.0460\pm0.0013$ & $-122.0\pm10.7$ & $4.00\pm0.06$ & 4.3 & 3.22 & 2.32 & 15.86 \\
18846 & $2.29\pm0.04$ & $-9.2\pm47.0$ & $61.6\pm1.8$ & $0.0619\pm0.0009$ & $26.8\pm13.6$ & $2.96\pm0.04$ & 5.9 & 1.55 & 1.02 & 9.27 \\
18976 & $0.79\pm0.02$ & $3.2\pm47.7$ & $41.8\pm0.0$ & $0.2841\pm0.0004$ & $49.4\pm5.3$ & $1.52\pm0.00$ & 8.3 & 1.26 & 1.06 & -1.30 \\
19342 & $1.03\pm0.02$ & $-18.4\pm32.8$ & $58.2\pm1.3$ & $0.0262\pm0.0001$ & $225.8\pm7.7$ & $4.22\pm0.01$ & 1.3 & 0.86 & 0.76 & -2.53 \\
19606 & $1.37\pm0.02$ & $-5.1\pm50.3$ & $56.3\pm1.4$ & $0.0290\pm0.0002$ & $49.9\pm10.4$ & $4.06\pm0.02$ & 1.5 & 1.16 & 1.13 & -2.26 \\
20961 & $0.51\pm0.01$ & $-22.2\pm45.8$ & $47.2\pm0.4$ & $0.0296\pm0.0001$ & $39.2\pm6.3$ & $3.60\pm0.01$ & 0.6 & 1.31 & 1.32 & -3.60 \\
22251 & $2.65\pm0.05$ & $-7.2\pm47.4$ & $51.5\pm1.0$ & $0.2919\pm0.0001$ & $22.4\pm9.2$ & $1.36\pm0.00$ & 24.5 & 1.73 & 1.66 & -2.33 \\
10009693 & $0.25\pm0.01$ & $14.0\pm0.0$ & $62.1\pm0.6$ & $0.1024\pm0.0006$ & $171.9\pm0.2$ & $5.89\pm0.05$ & 1.1 & 0.68 & 0.63 & -1.68 \\
10013609 & $1.16\pm0.02$ & $-20.6\pm43.3$ & $43.8\pm0.2$ & $0.3577\pm0.0017$ & $56.2\pm16.8$ & $2.08\pm0.01$ & 13.6 & 2.32 & 0.99 & 32.53 \\
10013620 & $0.88\pm0.03$ & $-8.2\pm49.2$ & $64.7\pm1.8$ & $0.0765\pm0.0002$ & $-90.1\pm7.8$ & $2.06\pm0.00$ & 2.0 & 0.95 & 0.89 & -2.36 \\
10013704 & $2.04\pm0.03$ & $-45.2\pm50.7$ & $62.6\pm2.3$ & $0.1323\pm0.0008$ & $-42.1\pm11.0$ & $2.29\pm0.04$ & 9.5 & 1.53 & 1.00 & 7.32 \\
10013905 & $0.92\pm0.02$ & $-0.0\pm50.1$ & $52.0\pm1.1$ & $0.0797\pm0.0012$ & $121.5\pm9.5$ & $2.26\pm0.01$ & 2.6 & 1.20 & 0.96 & 2.21 \\
10015338 & $1.76\pm0.06$ & $-45.5\pm45.5$ & $86.0\pm3.3$ & $0.0724\pm0.0026$ & $-239.3\pm3.0$ & $5.06\pm0.08$ & 1.9 & 1.04 & 0.75 & 3.32 \\
10016374 & $1.76\pm0.04$ & $-29.1\pm49.5$ & $79.8\pm2.7$ & $0.0347\pm0.0008$ & $208.1\pm13.2$ & $2.92\pm0.00$ & 2.1 & 2.40 & 1.99 & 2.97 \\
10056849 & $0.95\pm0.02$ & $10.7\pm40.1$ & $51.9\pm0.7$ & $0.0364\pm0.0002$ & $-287.7\pm2.1$ & $3.73\pm0.02$ & 1.3 & 0.99 & 0.94 & -2.44 \\
\hline                  
\end{tabular}
\\{\bf Note:} (1) NIRSpec ID of the target; (2) amplitude of the narrow component; (3,4) velocity centroid and dispersion of the narrow Gaussian component with respect to the systemic redshift of the galaxy reported in \cite{Bunker:2023}; (5) broad-to-narrow Gaussian amplitude  ratio; (6) velocity centroid of the broad Gaussian component with respect to the systemic redshift of the galaxy; (7)  broad-to-narrow Gaussian velocity dispersion ratio; (8) signal-to-noise ratio of the integrated flux of the broad component; (9,10) reduced chi-squared  for the single- and double-Gaussian models, respectively; (11) log-difference between the two Bayesian evidences of the single- and double-Gaussian models.\\
\end{table}
\end{landscape}

\section{Selection bias}\label{sec:selection_bias}
\begin{figure}
\includegraphics[width=0.5\textwidth]{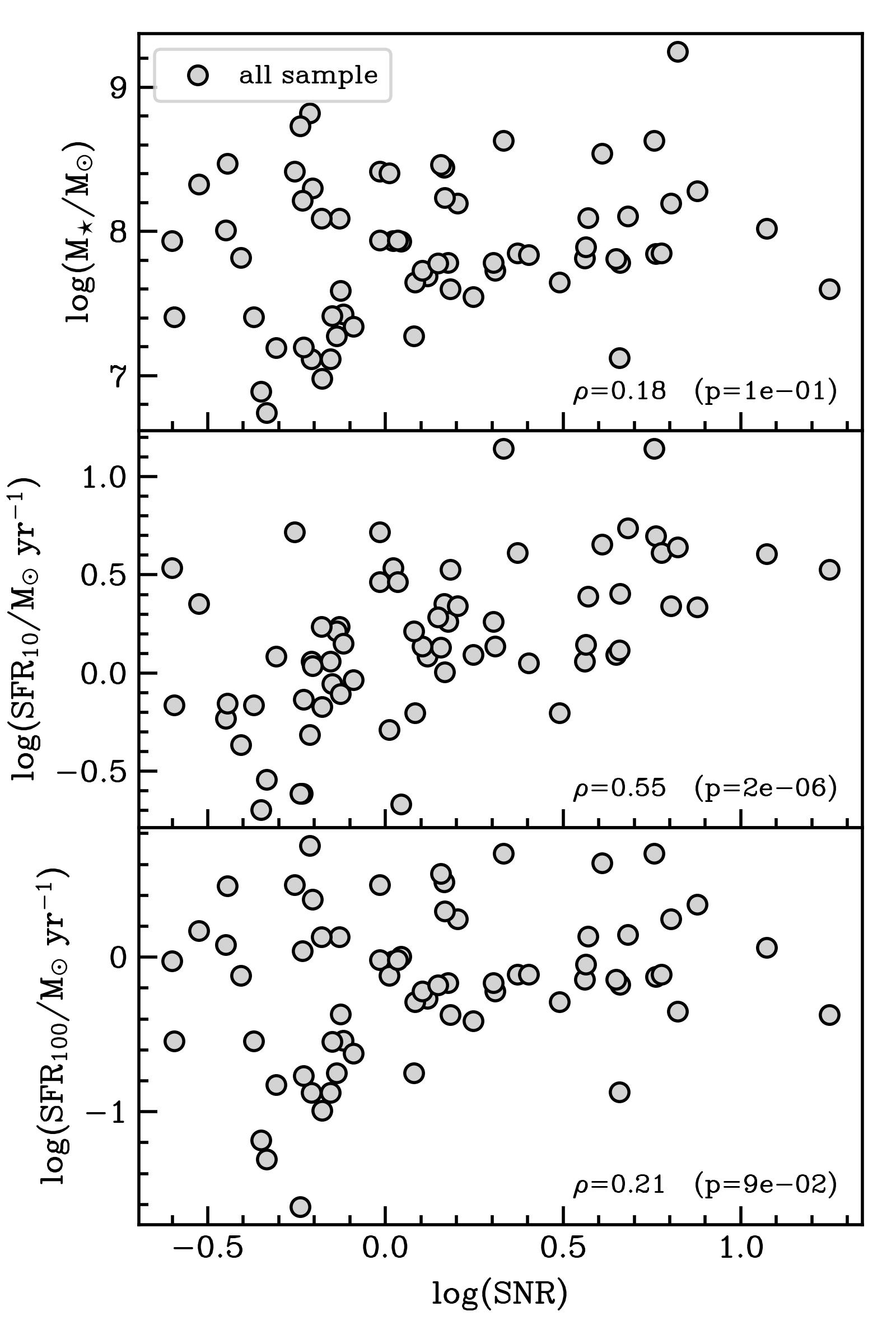}
\caption{Stellar mass, SFR (over the last 10 Myr), SFR$_{100}$ (over the last 100 Myr) of the JADES galaxies as a function of S/N of the nebular lines.  The Spearman rank correlation coefficient ($\rho$)  is reported in each plot, along with the corresponding
p-value.}
\label{fig:galaxy_vs_snr}
\end{figure}

A correlation between the incidence rate of outflows and the stellar mass and SFR may arise from the fact that more massive and starburst galaxies have stronger emission lines that increase the chance of detecting a broad component. To test this, we analysed whether the stellar mass and  SFR estimates in our sample scaled directly with the signal-to-noise ratio of the optical lines. 
The top and middle panels of Figure~\ref{fig:galaxy_vs_snr} show the distribution of stellar mass and SFR over the last 10~Myr, and they report the non-parametric
Spearman rank and p-value correlation parameter to test the hypothesis of a correlation between the parameters.
A value $>0.5$ of the Spearman rank combined with a
low p-value ($<0.05$) indicates a statistically significant
correlation between SFR and S/N. This suggests that the increase in the outflow incidence with SFR is primarily driven by the S/N.
 The bottom panel of Figure~\ref{fig:galaxy_vs_snr} reports the  SFR averaged over the last 100 Myr (SFR$_{100}$) as a function of the S/N. The estimated Spearman rank is much lower than 0.5 ($\sim$0.2), which indicates a weak correlation of SFR$_{100}$ with the S/N of the nebular lines. The same applies to the galaxy mass.
 Based on these results, we conclude that the observed correlation between outflow incidence and both \mstar\ and SFR$_{100}$ traces a genuine dependence on these quantities, while the correlation with the instantaneous (over 10 Myr) SFR is partly an effect of the S/N ratio of the lines.


\section{Outflow properties}\label{app:outflow_prop}
\begin{figure*}
\includegraphics[width=0.5\textwidth]{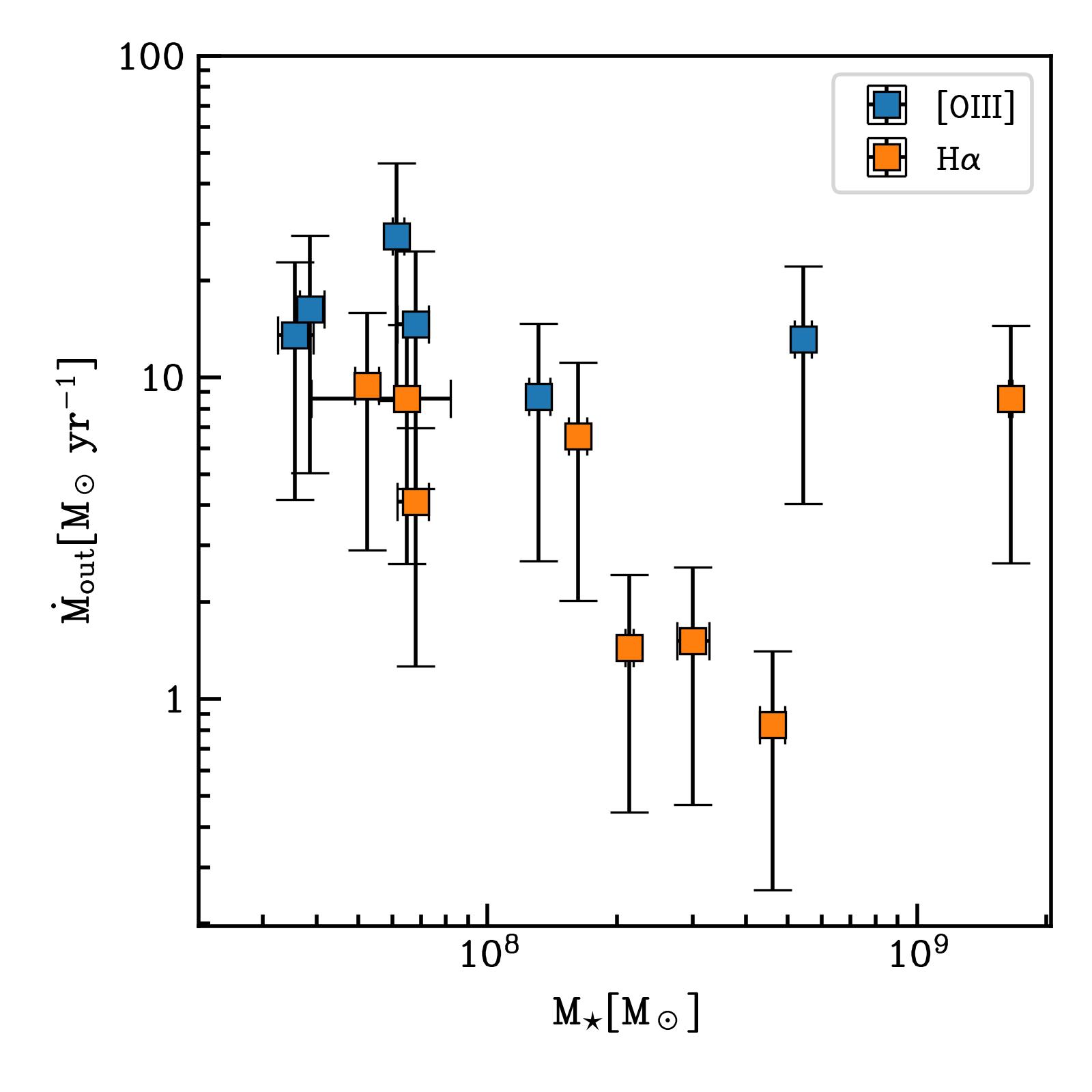}
\includegraphics[width=0.5\textwidth]{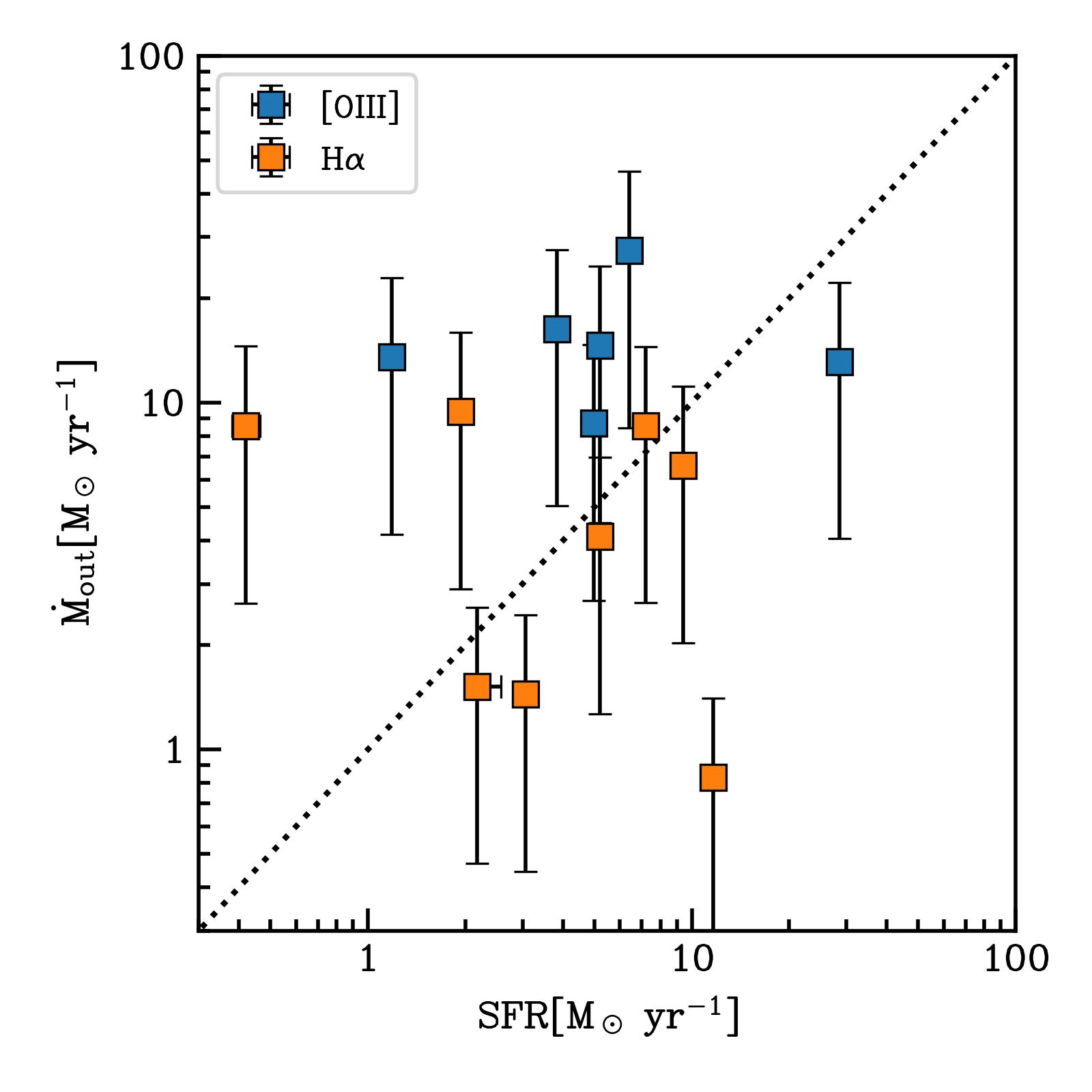}
\caption{Mass outflow rate as a function of stellar mass (left) and star formation rate (right). The dotted line in the right panel shows the 1:1 relation.}
\label{fig:mout_vs_mstarSFR}
\end{figure*}

Figure~\ref{fig:mout_vs_mstarSFR} reports the mass outflow rate estimates as a function of stellar mass and star formation rate. We do not find any clear correlation between mass outflow rate and galaxy properties.

\end{appendix} 

\end{document}